\documentclass[%
 aip,
 amsmath,amssymb,
 reprint,%
]{revtex4-2}

\usepackage{graphicx}
\usepackage{dcolumn}
\usepackage{bm}

\usepackage[utf8]{inputenc}
\usepackage[T1]{fontenc}
\usepackage{mathrsfs}
\usepackage{etoolbox}
\usepackage{bm}
\usepackage[uline]{hhtensor}
\usepackage{amsfonts}
\usepackage{xcolor}
\usepackage{siunitx}

\usepackage{draftwatermark}
\SetWatermarkText{\bf PREPRINT ~~~ PREPRINT ~~~ PREPRINT   ~~~ PREPRINT}
\SetWatermarkFontSize{0.8cm}
\SetWatermarkAngle{90}
\SetWatermarkHorCenter{20.5cm}
\SetWatermarkColor[gray]{0.75}

\makeatletter
\def\@email#1#2{%
 \endgroup
 \patchcmd{\titleblock@produce}
  {\frontmatter@RRAPformat}
  {\frontmatter@RRAPformat{\produce@RRAP{*#1\href{mailto:#2}{#2}}}\frontmatter@RRAPformat}
  {}{}
}%
\makeatother
\begin{document}

\preprint{AIP/123-QED}

\title{Modelling high-Mach-number rarefied crossflows past a flat plate using the maximum-entropy moment method}

\author{Stefano Boccelli}
\email{sboccell@uottawa.ca}
\affiliation{%
 Department of Mechanical Engineering, University of Ottawa\\
 161 Louis-Pasteur, ON K1N 6N5, Canada
}%

\author{Pietro Parodi}
\affiliation{%
  Department of Mathematics, KU Leuven\\
  Celestijnenlaan 200B, 3001 Leuven, Belgium
}%
\affiliation{%
  Aeronautics and Aerospace Department, von Karman Institute for Fluid Dynamics\\
  Waterloosesteenweg 72, B-1640 Sint-Genesius-Rode, Belgium
}%

\author{Thierry E. Magin}
\affiliation{%
  Aeronautics and Aerospace Department, von Karman Institute for Fluid Dynamics\\
  Waterloosesteenweg 72, B-1640 Sint-Genesius-Rode, Belgium
}%
\affiliation{%
  Aero-Thermo-Mechanics Department, Université libre de Bruxelles\\
  Avenue Franklin Roosevelt 50, 1050 Brussels, Belgium
}

\author{James G. McDonald}
\affiliation{%
 Department of Mechanical Engineering, University of Ottawa\\
 161 Louis-Pasteur, ON K1N 6N5, Canada
}%

\date{\today}

\begin{abstract}

The 10 and 14-moment maximum-entropy methods are applied to the study of high-Mach-number non-reacting crossflows past a flat plate at 
large degrees of rarefaction.
The moment solutions are compared to particle-based kinetic solutions, showing a varying degree of accuracy.
At a Knudsen number of 0.1, the 10-moment method is able to reproduce the shock layer, while it fails to predict the low-density wake region, 
due to the lack of a heat flux.
Conversely, the 14-moment method results in accurate predictions of both regions.
At a Knudsen number of 1, the 10-moment method produces unphysical results in both the shock layer and in the wake.
The 14-moment method also shows a reduced accuracy, but manages to predict a reasonable shock region, free of unphysical sub-shocks, 
and in qualitative agreement with the kinetic solution.
Accuracy is partially lost in the wake, where the 14-moment method predicts a thin unphysical high-density layer, concentrated on the 
centreline.
An analysis of the velocity distribution functions (VDF) indicates strongly non-Maxwellian shapes, and the presence of distinct particle
populations, in the wake, crossing each other at the centreline.
The particle-based and the 14-moment method VDFs are in qualitative agreement.
\end{abstract}

\maketitle


\section{Introduction}

The low-collisionality conditions encountered in high-altitude atmospheric flight\cite{lofthouse2007effects} and in laboratory vacuum equipment\cite{rebrov2001free}
are such that the translational modes of gas particles might realize strong degrees of thermodynamic non-equilibrium. 
Microscopically, translational non-equilibrium implies the presence of non-Maxwellian velocity distribution functions (VDFs).\cite{ferziger1972mathematical}
This has a number of macroscopic implications on the flow field.
For instance, in the rarefied regime one often encounters thick shock waves, pressure anisotropy, and a significant heat flux,
associated to the VDF asymmetry.
In such conditions, the traditional Navier-Stokes-Fourier (NSF) equations often fail to accurately reproduce 
the transport processes, such as the shear stress and heat conduction, while the Euler equations for gas dynamics neglect these
effects altogether.

The Knudsen number, $\mathrm{Kn} = \lambda/L$, expressed as the ratio of the collisional mean free path, $\lambda$, to a characteristic dimension of the problem, $L$,
offers a convenient measure of the expected degree of non-equilibrium, in a given flow.\cite{josyula2011review} 
For $\mathrm{Kn} > 0.001$, one leaves the continuum regime, 
where the Euler and the NSF equations of gas dynamics are known to be accurate, and enters the slip regime ($\mathrm{Kn}\approx 0.01$),
the transitional regime ($0.1 \lesssim \mathrm{Kn} \lesssim 10$), and eventually the free-molecular regime ($\mathrm{Kn} > 10$).
An accurate theoretical framework for studying flows in both continuum and rarefied conditions is the kinetic theory of gases,\cite{ferziger1972mathematical} 
where the phase-space evolution of the velocity distribution function is tracked.

Numerically, kinetic solutions are often obtained via the Direct Simulation Monte Carlo (DSMC) procedure,\cite{bird1994molecular}
where a reduced set of macro-particles is simulated:
these particles are advected through the domain, ballistically, taking into account solid-wall scattering events, 
and gas-phase collisions are performed among particles located within the same computational cell.
The computational cost of DSMC is low at high Knudsen numbers, but drastically increases as the translational regime is approached.
Moreover, the statistical formulation of this method introduces numerical noise, which may be undesirable in certain situations.
Another possibile approach consists in the discrete-velocity method (DVM), where one solves the kinetic equation on a phase-space grid, 
where both the physical space and the velocity space are discretized.\cite{mieussens2000discrete}
The DVM method is typically computationally expensive and memory-demanding, as it requires, in the general case, a six-dimensional grid.
It should be mentioned that a variety of models and numerical methods has been developed for treating the collision operator, 
for both Monte Carlo and DVM methods.\cite{aristov2019direct,kim2023critical,frezzotti2011solving}

Moment Methods are fluid-like formulations that extend the validity of traditional fluid models towards the rarefied regime.\cite{torrilhon2016modeling}
These formulations are derived directly from kinetic theory,\cite{ferziger1972mathematical}
and retain the capability to reproduce a range of non-equilibrium situations.
Moment methods are typically much more computationally affordable than the DVM or the DSMC methods, 
especially at intermediate Knudsen numbers, but are also less accurate in certain non-equilibrium scenarios.
Some widely employed moment methods include the Grad method\cite{grad1949kinetic} and its regularizations,\cite{struchtrup2003regularization,cai2014globally}
quadrature-based closures\cite{fox2009higher} and others.\cite{struchtrup2005macroscopic}
In this work, we consider the maximum-entropy moment method.\cite{muller1993extended,levermore1996moment}
Among its benefits, the maximum-entropy method is built on a non-perturbative representation of the VDF, that allows for large deviations 
from equilibrium.
Also, the resulting VDF is positive by construction, and the method can be shown to result in a system of globally-hyperbolic partial differential equations
for the time and space evolution of the moments of interest.
In the past, maximum-entropy-based formulations have been applied to a number of scenarios, including
polydisperse flows,\cite{forgues2019gaussian} rarefied supersonic jets,\cite{boccelli2023rarefied} hypersonic flows\cite{jayaraman2017multi} and 
charge transport in semiconductors.\cite{romano20022d}

In this work, we aim to apply the maximum-entropy moment method to high-supersonic (Mach 10) two-dimensional cross-flows of argon past a flat plate.
We analyze two different rarefaction conditions, considering the Knudsen numbers $\mathrm{Kn} = 0.1$ and $\mathrm{Kn}=1$.
We solve the 10 and 14-moment maximum-entropy methods and compare the results with particle-based solutions of the kinetic equation.
In this work, we focus on translational non-equilibrium, and neglect atomic processes such as chemical reactions and excitation/ionization.

In Section~\ref{sec:max-ent-formulation}, we briefly discuss the formulation of maximum-entropy moment methods, starting from the kinetic theory.
Section~\ref{sec:numerical-methods} discusses the numerical methods employed to solve the kinetic equation and the moment methods.
Section~\ref{sec:test-cases} discusses two different test cases, at $\mathrm{Kn}=0.1$ and $1$.
Finally, Section~\ref{sec:VDFs} analyzes the velocity distribution functions from the particle solution, at selected locations,
and compares them to the VDFs from the 14-moment model.
Conclusions are drawn in Section~\ref{sec:conclusions}.


\section{Maximum-entropy moment methods}\label{sec:max-ent-formulation}

In the kinetic theory of gases, one describes the phase-space evolution of the particle velocity distribution function (VDF), 
$f(\bm{x}, \bm{v}, t)$, via a kinetic equation.
For a non-reacting single-species gas, in absence of external forces, 
\begin{equation}\label{eq:kinetic-eq-BGK}
  \frac{\partial f}{\partial t} 
+ \bm{v} \cdot \frac{\partial f}{\partial \bm{x} } 
=
  \mathcal{C} 
= - \nu_c \left( f - \mathcal{M} \right) \, ,
\end{equation}

\noindent where $\bm{x}$ and $\bm{v}$ are the physical and velocity space respectively, and $t$ is the time.
In Eq.~\eqref{eq:kinetic-eq-BGK}, we employ the Bhatnagar-Gross-Krook (BGK) approximation for the collision operator,\cite{bhatnagar1954model} 
for simplicity, where the velocity-independent collision
frequency, $\nu_c$, is written as simple function of the Maxwellian mean free path, $\lambda$, and of the thermal velocity, $v^{\mathrm{th}}=\sqrt{8 p/\pi\rho}$,
where $p$ and $\rho$ are the hydrostatic pressure and the gas density respectively,
\begin{equation}\label{eq:collision-frequency}
  \nu_c = \frac{v^{\mathrm{th}}}{\lambda} = \frac{\sqrt{2} \, \rho \sigma}{m} \sqrt{\frac{8 p}{\pi \rho}} \, .
\end{equation}

\noindent Here, $m$ is the mass of argon atoms and $\sigma$ is a hard-sphere collisional cross-section, assumed to be constant for simplicity.

The BGK operator is known to be a radical approximation to the actual integro-differential Boltzmann collisional operator.
The latter should be employed for physically accurate computations. 
However, in this work we are concerned with comparing the kinetic and the moment methods. 
For this reason, we purposefully select the BGK operator, that allows for a direct comparison between all methods.
The BGK collision operator can in fact be implemented in a particle-based method, as discussed in previous 
studies.\cite{gallis2000application,macrossan2001particle}
Also, while chemical reactions and/or excitation/ionization processes are expected to play an important role at the considered Mach number, $\mathrm{M} = 10$, 
these phenomena are entirely neglected in this work, for the purpose of isolating only translational non-equilibrium effects.

The domain of the kinetic equation is the phase space, that encompasses the physical space and the particle velocity space.
In the most general case, the phase space is six-dimensional and obtaining a direct numerical solution is rarely feasible. 
In order to reduce this complexity, instead of solving the full kinetic equation, one can compute a finite set of equations for its 
statistical moments, $U_\phi$, defined as
\begin{equation}
  U_\phi = \iiint_{-\infty}^{+\infty} \phi(\bm{v}) f(\bm{v}) \, \mathrm{d} \bm{v}  \equiv \left< \phi(\bm{v} ) f(\bm{v}) \right> \, ,
\end{equation}

\noindent where $\phi(\bm{v})$ is a function of the particle velocity.
This formulation allows one to compute macroscopic quantities from the VDF.
For instance, the density, $\rho$, the average momentum density in the $i$-th direction, $\rho u_i$, and the pressure tensor of a gas, $P_{ij}$, read
\begin{equation}
  \rho = \left<m f\right> \ \ , \ \ \ \rho u_i = \left< m v_i f \right> \ \ , \ \ \ P_{ij} = \left< m c_i c_j f \right> \, ,
\end{equation}

\noindent where $u_i$ is the bulk velocity, and $c_i = v_i - u_i$ is the peculiar velocity.
Moments obtained about the peculiar velocity do not contain a convection component.
These only depend on the shape of the distribution function, and not on the bulk velocity.
These are referred to as central moments.
The hydrostatic pressure of the gas, $p$, is obtained from the trace of the pressure tensor, $p = P_{ii}/3$, where repeated indices 
imply summation.
The temperature, $T$, is obtained from the ideal gas law,
\begin{equation}
  p = n k_B T \, ,
\end{equation}

\noindent where $n = \rho / m$ is the number density and $k_B$ is the Boltzmann constant.
Higher-order central moments can also be computed. 
For instance, the heat flux tensor, $Q_{ijk}$, and the fourth-order tensor, $R_{ijkl}$, can be written as
\begin{equation}\label{eq:kinetic-definition-Qijk-Rijkl}
  Q_{ijk} = \left< m c_i c_j c_k f \right> \ \ , \ \ \ R_{ijkl} = \left< m c_i c_j c_k c_l f \right> \, .
\end{equation}

The contracted fourth-order moment, $R_{iijj}$, is proportional to the density and to the fourth-order power of the particle velocity, 
making it an indicator for the kurthosis of the distribution function.
The heat flux vector, $q_i$, can be obtained from the contraction of the heat flux tensor, 
\begin{equation}
  q_i = \tfrac{1}{2} Q_{ijj} \, ,
\end{equation}

\noindent where the multiplying factor, $1/2$, comes from the fact that, in Eq.~\eqref{eq:kinetic-definition-Qijk-Rijkl}, 
$Q_{ijk}$ was defined as \textit{twice} of the heat flux.
In this work, we simply refer to $Q_{ijk}$ as the heat flux tensor, without further mentioning the factor $1/2$.
Finally, the fifth-order tensor, $S_{ijklm}$, is obtained as
\begin{equation}
  S_{ijklm} = \left< m c_i c_j c_k c_l c_m f \right> \, .
\end{equation}

An evolution equation for a moment of interest, $U_\phi$, is obtained by first multiplying the full kinetic equation by the generating function, 
$\phi(\bm{v})$, and then by integrating over the velocity space.\cite{ferziger1972mathematical}
This approach can be used to generate a hierarchy of moment equations, written in balance-law form, 
\begin{equation}
  \frac{\partial \bm{U}}{\partial t}
+ \frac{\partial \bm{F}_x}{\partial x}
+ \frac{\partial \bm{F}_y}{\partial y}
+ \frac{\partial \bm{F}_z}{\partial z}
= 
  \bm{S} \, ,
\end{equation}

\noindent where $\bm{U}$ is a vector collecting all moments of interest, 
$\bm{F}_i$ are the vectors of convected fluxes in the $i$-th direction and $\bm{S}$ is a vector of collisional source terms.
This constitutes an infinite hierarchy of moment equations, that requires a closure.
In the maximum-entropy method, a closure is obtained by prescribing that, for a given set of moments of interest, $\bm{U}$, 
the distribution function is the one that maximizes the statistical entropy, 
\begin{equation}
  \mathcal{S} = - k_B \iiint_{-\infty}^{\infty} f \ln{f/\chi} \, \mathrm{d} \bm{v} \, ,
\end{equation}

\noindent where $\chi$ is a normalizing constant.
It can be shown that the VDF that maximizes this entropy takes the form\cite{dreyer1987maximisation} 
\begin{equation}
  f_{\mathrm{ME}} = \exp \left( \bm{\alpha}^\intercal \bm{\Phi}(\bm{v}) \right) \, ,
\end{equation}

\noindent where $\bm{\alpha}$ is a vector of coefficients and $\bm{\Phi}(\bm{v})$ is the vector collecting all generating functions, $\phi(\bm{v})$, 
associated with $\bm{U}$.
The system of partial differential equations resulting from the maximum-entropy closure can be shown to be globally hyperbolic, 
as the flux Jacobian has real eigenvalues for all gas states.
For the full derivations and additional details, the reader is referred to the works by Levermore,\cite{levermore1996moment}
Dreyer\cite{dreyer1987maximisation} and M{\"u}ller \& Ruggeri.\cite{muller1993extended}
Here, we only summarize a few members of the maximum-entropy family of moment methods, originating from different choices of the moments of interest.


\subsection{The Euler equations}\label{sec:theory-5-moment-method}

The simplest maximum-entropy moment system is second-order in the particle velocity and is associated with a VDF in the form
\begin{equation}
  f_5 = \exp \left(\alpha_0 + \alpha_i v_i + \alpha_2 v^2 \right) \, ,
\end{equation}

\noindent that is a function of five independent parameters (in three physical dimensions, for $i=(x,y,z)$) and coincides with a Maxwellian distribution,
\begin{equation}
  \mathcal{M} = \frac{\rho}{m} \left( \frac{\rho}{2 \pi p} \right)^{3/2} \exp \left[ - \frac{\rho}{2 p} \left(\bm{v} - \bm{u}\right)^2 \right] \, .
\end{equation}

Being isotropic (in the frame of the bulk velocity, $\bm{u}$), this VDF is associated with a zero heat flux and no shear stress.
The resulting vector of moments of interest, written using primitive variables, is
\begin{equation}
  \bm{W}_5 = \left( \rho, u_x, u_y, u_z, p \right) \, .
\end{equation}

\noindent and the simplest maximum-entropy system happens to coincide with the Euler equations of gas dynamics.
This model is introduced only for completeness, and is not discussed further in this work.


\subsection{The Gaussian 10-moment system}\label{sec:theory-10-moment-method}

A more complex maximum-entropy moment system, still of second order in the velocity, is obtained from a VDF in the form
\begin{equation}
  f_{10} = \exp \left(\alpha_0 + \alpha_i v_i + \alpha_{ij} v_i v_j \right) \, .
\end{equation}

\noindent This VDF depends on ten parameters and coincides with the anisotropic Gaussian distribution,
\begin{equation}
  \mathcal{G} = \frac{\rho}{m} \frac{1}{(2 \pi)^{3/2}} \frac{1}{(\det \Theta)^{1/2}} \exp\left[ -\frac{1}{2} \Theta_{ij}^{-1} c_i c_j \right] \, ,
\end{equation}

\noindent where $\Theta_{ij} = P_{ij}/\rho$ is a symmetric tensor.
The iso-surfaces of this distribution are ellipsoids, oriented along the principal axes of $\Theta_{ij}$.
This symmetry makes it unable to support a heat flux.
However, as it can be rotated in velocity space, this distribution can support non-zero shear stresses. 
The vector of the ten primitive moments of interest is
\begin{equation}
  \bm{W}_{10} = \left( \rho, u_x, u_y, u_z, P_{xx}, P_{xy}, P_{xz}, P_{yy}, P_{yz}, P_{zz} \right) \, ,
\end{equation}

\noindent and the 10-moment maximum-entropy system reads, in index notation,
\begin{subequations}
\begin{equation}
  \frac{\partial \rho}{\partial t}
+ \frac{\partial \rho u_k}{\partial x_k}
= 0 \, ,
\end{equation}
\begin{equation}
  \frac{\partial \rho u_i}{\partial t}
+ \frac{\partial}{\partial x_k} \left( \rho u_i u_k + P_{ik} \right)
= 0 \, ,
\end{equation}
\begin{multline}\label{eq:10mom-Pij-eq}
  \frac{\partial}{\partial t} \left( \rho u_i u_j + P_{ij} \right)
+ \frac{\partial}{\partial x_k} \left( \rho u_i u_j u_k + u_i P_{jk} \right. \\
   \left. + u_j P_{ik} + u_k P_{ij} \right)
= -\nu_c \left( P_{ij} - p \, \delta_{ij}\right) \, .
\end{multline}
\end{subequations}

Notice that the trace of the momentum-flux tensor, $\rho u_i u_j + P_{ij}$, is (twice of) the energy density of the gas, $\rho E$.
Therefore, the energy equation is automatically embedded in the 10-moment system.
The collisional source term, at the right-hand side, relaxes the pressure tensor, $P_{ij}$, to its isotropic version, $p\,\delta_{ij}$.
Considering the $x$ direction, the flux Jacobian has six repeated wave speeds, equal to the bulk velocity, $u_x$, and four 
faster wave speeds, that include the $x$-directed characteristic velocity, $c_{xx} = \sqrt{P_{xx}/\rho}$, that is related to the speed of sound.
We have:
\begin{equation}
  \lambda_x = (u_x \left\{ 6\times \right\} \ , \ \ u_x \pm \sqrt{P_{xx}/\rho} \ ,  \ \  u_x \pm \sqrt{3 P_{xx}/\rho} ) \, .
\end{equation}

\noindent The wave speeds along $y$ and $z$ are analogous.
For a further discussion, and for some applications of this system to one and two-dimensional test cases, 
the reader is referred to McDonald and Groth, 2005.\cite{mcdonald2005numerical}


\subsection{The 14-moment system}\label{sec:theory-14mom-method}

The 14-moment system constitutes the simplest fourth-order maximum-entropy method, and is associated with a 14-parameter VDF in the form
\begin{equation}\label{eq:f14-equation}
  f_{14} = \exp \left( \alpha_0 + \alpha_i v_i + \alpha_{ij} v_i v_j + \alpha_{ijj} v_i v^2 + \alpha_{iijj} v^4 \right) \, .
\end{equation}

At equilibrium, this distribution recovers the Maxwellian distribution, but can also realize a range of non-equilibrium
shapes, including the Gaussian, bi-modals, the Druyvesteyn distribution, and others.\cite{boccelli202014}
For an overview of the 14-moment distributions associated with various non-equilibrium moment states, see Boccelli et al, 2023.\cite{boccelli2023gallery}

The 14-moment VDF is associated with the following primitive variables,
\begin{equation}
  \bm{W}_{14} = \left( \rho, u_i, P_{ij}, Q_{ijj}, R_{iijj} \right) \, .
\end{equation}

\noindent Notice that the 14-moment model does not employ, as moments of interest, the full tensors $Q_{ijk}$ and $R_{ijlk}$, but only
their contractions, namely the heat flux \textit{vector}, $Q_{ijj}$, and the scalar quantity, $R_{iijj}$.
The 14-moment system of equations reads,\cite{mcdonald2013affordable}
\begin{widetext}
\begin{subequations}
\begin{equation}
    \frac{\partial}{\partial t} \rho 
    + \frac{\partial}{\partial x_i} \left( \rho u_i \right) = 0  \, ,
\end{equation}
\begin{equation}
    \frac{\partial}{\partial t} \left( \rho u_i \right)
    + \frac{\partial}{\partial x_j} \left( \rho u_i u_j + P_{ij} \right) = 0 \, ,
\end{equation}
\begin{equation}
    \frac{\partial}{\partial t} \left( \rho u_i u_j + P_{ij} \right)
    + \frac{\partial}{\partial x_k} \left( \rho u_i u_j u_k + u_i P_{jk} + u_j P_{ik} 
     + u_k P_{ij}  + Q_{ijk} \right) = -\nu_c \left(P_{ij} - p\,\delta_{ij} \right) \, ,
\end{equation}
\begin{multline}
    \frac{\partial}{\partial t} \left( \rho u_i u_j u_j + u_i P_{jj} + 2 u_j P_{ij} + Q_{ijj} \right)
    +\frac{\partial}{\partial x_k} \left(\rho u_i u_k u_j u_j  + u_i u_k P_{jj}
     + 2 u_i u_j P_{jk} + 2 u_j u_k P_{ij} + u_j u_j P_{ik}  \right. \\
    \left.  + u_i Q_{kjj} + u_k Q_{ijj} + 2 u_j Q_{ijk} + R_{ikjj} \right) 
= -\nu_c \left[ 2 u_j \left(P_{ij} - p\, \delta_{ij} \right) + Q_{ijj} \right]  \, ,
\end{multline}
\begin{multline}
    \frac{\partial}{\partial t} \left( \rho u_i u_i u_j u_j + 2 u_i u_i P_{jj} + 4 u_i u_j P_{ij} + 4 u_i Q_{ijj} + R_{iijj} \right) \\
    + \frac{\partial}{\partial x_k} \left( \rho u_k u_i u_i u_j u_j + 2 u_k u_i u_i P_{jj} + 4 u_i u_i u_j P_{jk}+ 4 u_i u_j u_k P_{ij}  + 2 u_i u_i Q_{kjj}  + 4 u_i u_k Q_{ijj} \right. \\
    \left. + 4 u_i u_j Q_{ijk} + 4 u_i R_{ikjj} + u_k R_{iijj} + S_{kiijj} \right) 
= -\nu_c \left[ 4 u_i u_j \left( P_{ij} - p\, \delta_{ij} \right) + 4 u_i Q_{ijj} + R_{iijj} - \frac{15 p^2}{\rho} \right] \, ,
\end{multline}
\end{subequations}
\end{widetext}

\noindent where the collisional right-hand-sides are obtained by subtracting the respective moment of interest, $\bm{U}$, from its equilibrium counterpart.
At equilibrium, the odd-order moments of the particle velocity (such as the heat flux) are zero, while 
the equilibrium value for the contracted fourth-order moment is $R_{iijj}^{\mathrm{eq}} = 15 p^2/\rho$. 

The convected fluxes contain some moments that do not appear in the vector of moments of interest, $\bm{U}$.
These moments are $Q_{ijk}$, $R_{ikjj}$ and $S_{kiijj}$.
In principle, one should do the following:
\begin{itemize}
  \item For every grid cell and at every time step of a numerical simulation, one should take the gas state, $\bm{U}$, 
        and compute the associated maximum-entropy distribution function, $f_{14}$, by solving a (numerical) optimization problem.\cite{schaerer2017efficient,boccelli2023gallery} 
  \item Once $f_{14}$ is computed, the closing moments can be obtained by direct integration in velocity space.
\end{itemize}

However, this procedure requires a substantial computational effort, which makes it challenging to compute anything but one-dimensional solutions.
An efficient solution of the entropy maximization problem is an active area of research, often addressed with the use of hardware acceleration.\cite{schaerer2017efficient,zheng2023stabilizing}
In this work, we employ a different approach, developed by McDonald and Torrilhon, 2013.\cite{mcdonald2013affordable}
Rather than maximizing the entropy numerically, we employ instead pre-computed analytical approximations of the closing moments, 
obtained as interpolative approximations of the maximum-entropy solution.
\footnote{Notice that a possible strategy to assess the accuracy of this approximation, in the context of the present simulations, 
could consist in computing the exact maximum-entropy closing moments, from a converged approximated solution, and comparing them to the approximation
itself.}
The expression of the closing fluxes is reported in the appendix, and the reader is referred to the original publication for a detailed discussion.
This approach makes the solution of the 14-moment method affordable.

Notice that the simpler maximum-entropy formulations (the 5 and 10-moment systems) do not present this difficulty.
Indeed, the simple shape of the VDF associated to these systems (respectively a Maxwellian or a Gaussian) permits to easily find 
an analytical expression for the closing moments.

From kinetic considerations, one can see that the 14 moments of interest are subject to both physical and mathematical limitations.
First of all, one cannot select independently all the moments, and must instead respect the inequality 
\begin{equation}
  R_{iijj} \ge Q_{kii} (P^{-1})_{kl} Q_{ljj} + \frac{P_{ii} P_{jj}}{\rho} \, ,
\end{equation}

\noindent This inequality stems from the non-negativity of the VDF, and is denoted as the physical realizability condition, 
in moment space.\cite{mcdonald2013affordable,boccelli2023gallery}
Practically speaking, this inequality does not constitute a problem: 
if the solution is initialized within the physical realizability boundary, the numerical solution automatically remains inside it, during a simulation.
Moreover, collisions tend to push the solution away from the realizability boundary, and towards equilibrium.

Second of all, Junk\cite{junk2002maximum} has shown the existance of a zero-measure region of moment space, located at
\begin{equation}
  Q_{ijj} = 0 \ , \ \ R_{iijj} \ge \frac{2 P_{ji} P_{ij} + P_{ii} P_{jj}}{\rho} \, ,
\end{equation}

\noindent where the entropy maximization problem does not admit a maximum, but only an extremum.
On that region, the flux Jacobian shows a singularity, that manifests as infinitely fast wave speeds.
While this poses some numerical difficulties, it should be noted that the presence of fast wave speeds opens the door to the possibility of 
reproducing infinitely smooth shock profiles,\cite{mcdonald2013affordable} even at high Mach numbers.
This is exactly what we aim at achieving in the present work.
In contrast, other moment methods are known to be affected by the presence of unphysical sub-shocks,\cite{taniguchi2018sub} 
whose origin is purely mathematical.


\section{Numerical methods}\label{sec:numerical-methods}

The kinetic equation, Eq.~\eqref{eq:kinetic-eq-BGK}, is solved in this work with a particle-based numerical method.\cite{bird1994molecular}
A set of numerical particles is injected from the boundaries and is advected across the domain.
Collisions are performed among particles belonging to the same grid cell.
In the present work, an unstructured grid of triangles is employed, and the total number of simulated particles, at steady state, is approximately 
ten millions for the test case of Section~\ref{sec:Kn-0p1-simulations} and four millions for the case of Section~\ref{sec:Kn-1-simulations}.
BGK collisions are implemented into the particle scheme as follows:
for each cell in the computational domain,
\begin{enumerate}
  \item The local density, average velocity, and temperature are computed from the available particles, and the local collision frequency, $\nu_c$, is
        obtained from Eq.~\eqref{eq:collision-frequency};
  \item Each particle in the cell is tested for collisions. A collision happens if a uniformly distributed random number, 
        $\mathcal{R}\in [0,1]$, satisfies the relation 
        \begin{equation}
          \mathcal{R} < 1 - \exp(- \nu_c \Delta t) \, ,
        \end{equation}
        \noindent where $\Delta t$ is the computational time step; 
  \item If the collision happens, the particle velocities are re-sampled from a Maxwellian distribution at the local density, velocity and temperature.
\end{enumerate}

\noindent Notice that this formulation conserves exactly the mass, but only conserves the momentum and energy statistically.
In the present simulations, this does not constitute an issue, given the open-boundary nature of the setup.
In the simulations, the time step is selected such as to resolve the local collision frequency, and the cells are spatially adapted in order to resolve the
local mean free path.
This algorithm was implemented and solved with the PANTERA Particle-in-Cell/Direct Simulation Monte Carlo (PIC/DSMC) solver.\cite{parodi2021pic}
Additional information on the time-convergence of the particle solutions is given in Appendix~\ref{sec:convergence-history}.

The solution of the maximum-entropy moment systems is obtained with a cell-centred finite-volume solver, on a Cartesian grid.
A first-order forward Euler time-explicit method is adopted, marching in time until convergence, as shown in Appendix~\ref{sec:convergence-history}.
Rusanov (local Lax-Friedrichs) numerical fluxes are employed to approximate the interface fluxes.\cite{toro2020low}
It should be noted that little information is presently available about the wave structure of the 14-moment maximum-entropy system.
Therefore, numerical schemes that require a limited knowledge of the wave speeds and the eigenvectors---such as central schemes with artificial dissipation---are preferred.
The maximum and minimum wave speeds of a PDE system are necessary to estimate the Courant number, and to compute the Rusanov flux functions.
An exact computation of the wave speeds would require to build explicitly the 14-by-14 flux Jacobian matrix, and compute its eigenvalues numerically.
This is computationally expensive. 
Therefore, we employ here previously developed approximated formulas for the maximum and minimum wave speeds.\cite{boccelli2023rarefied}

To speed-up the simulations, the fluid solutions are first converged in time on a coarse grid, using first-order spacial accuracy. 
Once a steady state is reached, the computation is restarted with second-order spacial accuracy (MUSCL method,\cite{van1979towards} with symmetric van Albada slope limiter\cite{toro2013riemann}) on progressively finer grids, until spacial convergence is obtained. 
This allows us to avoid computing the full transitory on a finer grid.
For the 14-moment system, the transitory happens to be the most challenging part, due to the presence of fast-moving waves.
As some of these waves are associated with the solution crossing the Junk subspace, the transitory also happens to be the less robust part of the simulation.
Starting from a coarse grid and working our way up also helps on this side.
Another helpful strategy could consist in starting the 14-moment solution from an already-converged lower-order solution, such as the 10-moment solution;
this approach was not tested in the present work and is suggested as a future development.
Further computational acceleration is here obtained by parallelizing the solution of the moment equations on an A100 NVIDIA GPU (CUDA Fortran implementation).

During the simulations, the 14-moment solution is artificially restrained from approaching the Junk line too closely.
This procedure is described in McDonald and Torrilhon, 2013,\cite{mcdonald2013affordable} and is obtained by enforcing a lower value on a parameter of the approximated closure, 
$\sigma_\mathrm{lim}$, defined in the said reference.
Here, $\sigma_\mathrm{lim}$ is not to be confused with the collisional cross-section.
A value of $\sigma_\mathrm{lim} = 10^{-4}$ is employed in this work and proved to be sufficient.
An analysis of the effect of $\sigma_{\mathrm{lim}}$ on the solution and on the computational cost, 
is provided in Boccelli et al, 2023.\cite{boccelli2023rarefied}
This reference also includes further implementation notes, as well as one-dimensional shock-wave problems and two-dimensional supersonic-jet test-cases for the 14-moment equations.

Zero-gradient boundary conditions (BCs) are imposed on the open boundaries of the simulation domain, and symmetry BCs are imposed on the $y=0$ axis.
The maximum-entropy systems are solved with the Hyper2D solver.\cite{hyper2Dgithub}
Specular-reflection boundary conditions are employed for the solid wall.
For the particle method, this implies that the velocity component perpendicular to the wall is reflected as a result of a 
particle-surface collision event.
In the fluid methods, the wall boundary conditions are implemented via ghost cells:
the ghost-cell value is adapted in time, before the computation of numerical fluxes, by copying the solution of the internal adjacent cell, 
but reflecting the moments that are proportional to odd-powers of the 
particle velocity component normal to the wall, $v_x$.
For the 14-moment system, these are $u_x$, $P_{xy}$, $P_{xz}$ and $Q_{xii}$.
Notice that, for an increased physical accuracy, one might want to employ diffusive-wall boundary conditions, instead of reflective walls.
This is suggested as a future work activity.
The solid wall is assumed to be 1 cm thick in the particle simulations and is infinitely thin in the fluid simulations. 
This thickness does not have an appreciable impact in the present simulations, 
being negligible with respect to both the considered mean free paths and to the transverse dimension of the flat plate.


\section{Test cases}\label{sec:test-cases}

\begin{figure}[htpb!]
  \centering
  \includegraphics[width=\columnwidth]{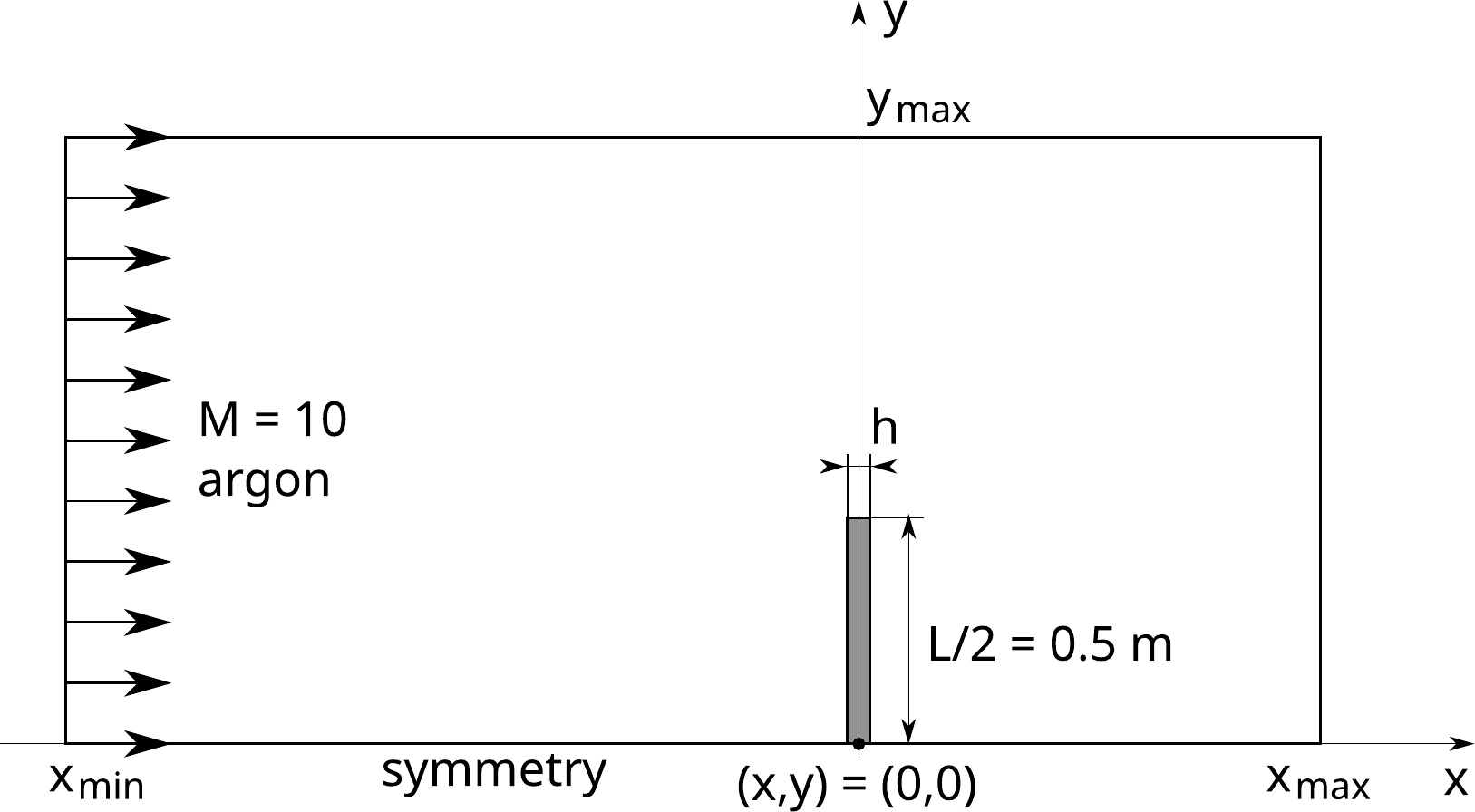}
  \caption{Geometry of the computational domain.}
  \label{fig:domain-scheme}
\end{figure}

In this work we study a two-dimensional crossflow of non-reacting argon gas past a flat plate.
The computational domain is sketched in Figure~\ref{fig:domain-scheme}.
Only half of the domain is simulated, and symmetry boundary conditions are imposed on the $y=0$ axis.
The free-stream Mach number is set to $\mathrm{M}=10$ for all considered simulations, with a free-stream temperature $T_{\mathrm{FS}}=300~\si{K}$.
Two different degrees of rarefaction are simulated, by varying the free-stream density.
The densities $\rho_{\mathrm{FS}} = 4.69\times{10^{-6}}~\si{kg/m^3}$ and  $\rho_{\mathrm{FS}} = 4.69\times{10^{-7}}~\si{kg/m^3}$ were considered, 
corresponding to the Knudsen numbers $\mathrm{Kn} = 0.1$ and $\mathrm{Kn} = 1$ respectively, based on a flat plate length $L = 1~\si{m}$,
\begin{equation}
  \mathrm{Kn} 
= \frac{\lambda}{L} 
= \frac{m}{\sqrt{2} \, \rho \sigma L} \, ,
\end{equation}

\noindent where $m = 6.6337\times10^{-26}~\si{kg}$ is the argon particle mass, 
and where $\sigma = 1\times 10^{-19}~\si{m^2}$ is a constant collision cross-section.
All other thermodynamic quantities characterizing the free stream are selected from a Maxwellian distribution.
The size of the computational domain is selected individually for each simulation.
Lower-collisionality simulations generally require larger domains, as the shock wave thickens.
The largest domain is employed for the $\mathrm{Kn} = 1.0$ simulations, with:
$(x_\mathrm{min},x_\mathrm{max}) = (-6, 3)~\si{m}$, and $(y_\mathrm{min},y_\mathrm{max}) = (0, 3)~\si{m}$.
For the $\mathrm{Kn}=0.1$ simulations, the shock is more localized, and the domain was reduced, longitudinally, to
$(x_\mathrm{min},x_\mathrm{max}) = (-3, 3)~\si{m}$.
The figures reported in this section only show a meaningful subregion of the full computational domains.

At the considered degree of rarefaction, we expect the solution to reach a stationary state, both in the shock region and in the wake.
This is confirmed by the result of our numerical simulations.
For a discussion of unsteady effects in blunt-body wakes and of the von Kármán street, 
at Knudsen numbers up to Kn=0.03, we refer the reader to Rovenskaya, 2022.\cite{rovenskaya2022subsonic}


\subsection{Kn = 0.1 simulations}\label{sec:Kn-0p1-simulations}

\begin{figure*}[htpb!]
  \centering
  \includegraphics[width=\textwidth]{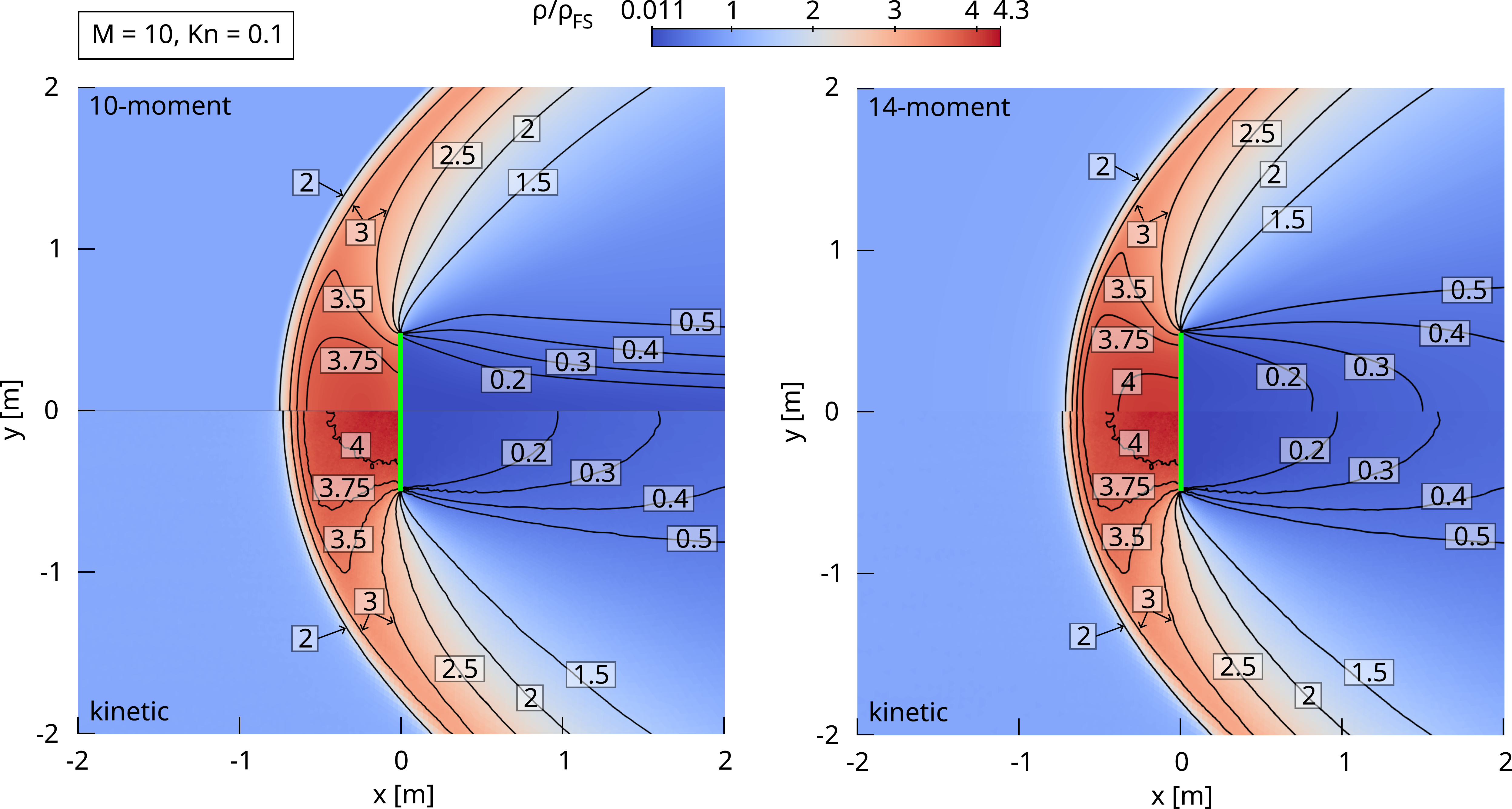}
  \caption{Density field, scaled with the free-stream density. Kinetic simulations (Bottom half of the plots, for $y<0$) compared to the 10-moment (Top-Left) and 14-moment (Top-Right) maximum-entropy simulations, at a Knudsen number Kn = 0.1.}
  \label{fig:simulations-Kn0p1-2Dplots}
\end{figure*}

\begin{figure*}[htpb!]
  \centering
  \includegraphics[width=\textwidth]{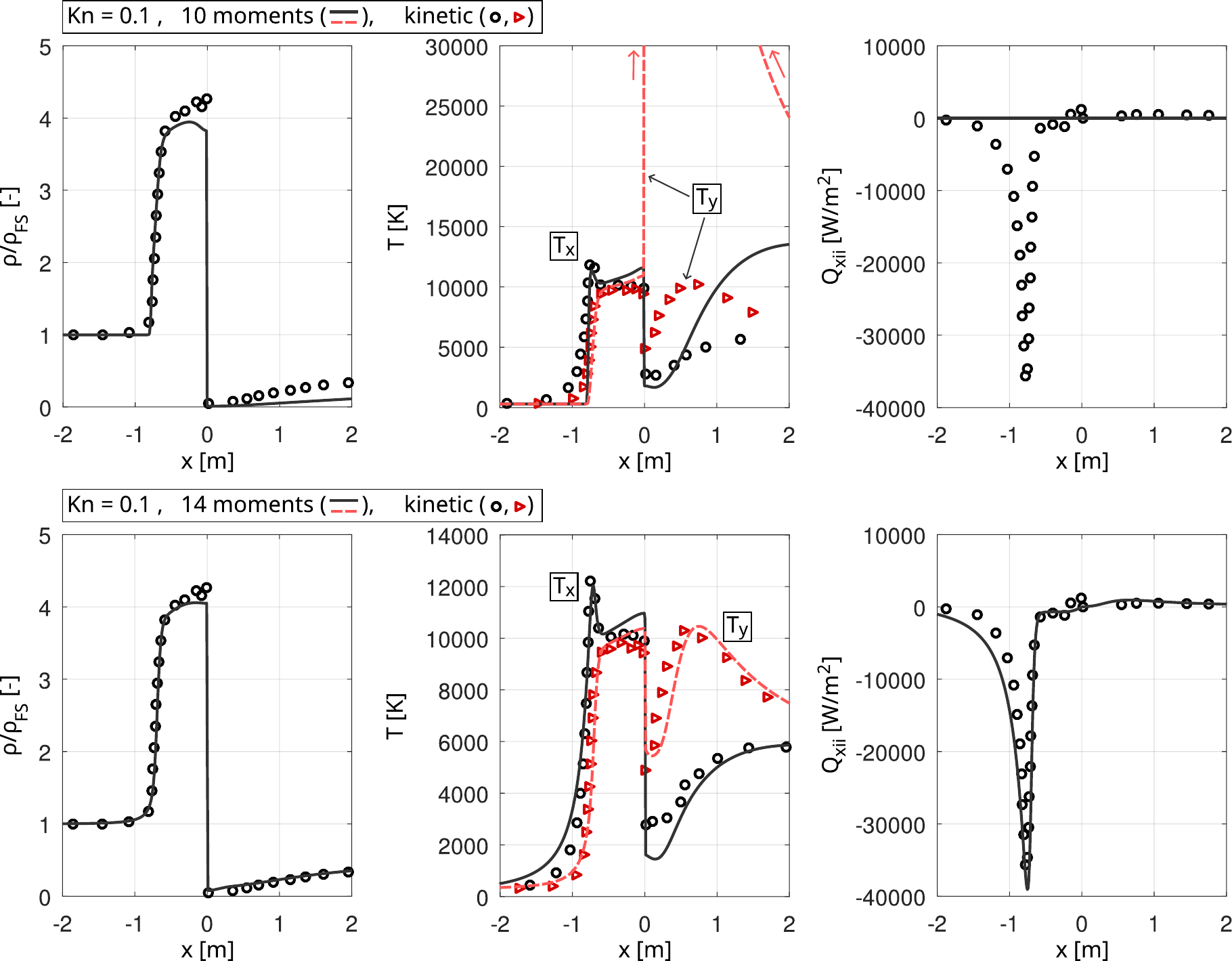}
  \caption{Selected moments along the symmetry line, for the Kn = 0.1 simulations. 
           Kinetic solution (symbols), 10-moment solution (lines, Top boxes) and 14-moment solution (lines, Bottom boxes).
           Positions with $x < 0$ correspond to the stagnation line, while $x > 0$ denotes the wake.}
  \label{fig:simulations-stagline-Kn0p1}
\end{figure*}

Figure~\ref{fig:simulations-Kn0p1-2Dplots} shows the two-dimensional density profiles for the kinetic and moment simulations, 
at $\mathrm{Kn} = 0.1$,
while Fig.~\ref{fig:simulations-stagline-Kn0p1} shows some selected fields, extracted along the symmetry plane, $y = 0$.
Considering the kinetic simulations (Bottom-half of Fig.~\ref{fig:simulations-Kn0p1-2Dplots}, and symbols in Fig.~\ref{fig:simulations-stagline-Kn0p1}),
one can first of all appreciate the finite thickness of the shock wave, that is of the order of the mean free path (in the present simulation, $\lambda = L \cdot \mathrm{Kn} = 0.1~\si{m}$). 

Inside the shock wave, the longitudinal velocity of the gas particles is converted into a longitudinal temperature, $T_{xx}$.
The perpendicular temperature, $T_{yy}$, lags behind, showing a strong temperature anisotropy, and $T_{xx}$ overshoots $T_{yy}$.
Eventually, collisions bring the gas state towards equilibrium, and the temperatures relax towards each other,\cite{muntz1969molecular}
although, in the present conditions, collisionality is not sufficient to reach a complete equilibrium in the post-shock region.
Behind the flat plate, under the combined effect of thermal diffusion and collisionality, the low-pressure wake is gradually replenished.
The noise showed by the particle simulation can be reduced by increasing the number of simulated particles, 
or by averaging the solution over a larger number of time steps.
In the present work, the level of noise does not constitute an issue for our analysis.

The 10-moment solution can be seen to reproduce the kinetic solution to a reasonable accuracy in the shock layer, in terms of the shock position, while 
the maximum density is slightly underpredicted.
A close inspection of the temperature field (Fig.~\ref{fig:simulations-stagline-Kn0p1}-Top-Centre) shows that 
the 10-moment method is not able to reproduce a smooth shock profile, and predicts instead a discontinuity, approximately at position $x = -0.8~\si{m}$.
While particularly evident in the temperature field, the discontinuity can also be appreciated from a close inspection of the density
field (Fig.~\ref{fig:simulations-stagline-Kn0p1}-Top-Left).
This is an expected behavior, since, at the considered Mach number, the maximum wave speed of the 10-moment system is not sufficiently large
to carry information upstream.

Remarkably, the 10-moment method manages to recover the overshoot of $T_{yy}$, in the immediate post-shock region, to a reasonable accuracy.
However, the 10-moment method fails in the wake, where no re-filling is predicted.
Instead, the 10-moment method appears to reproduce a fairly stable low-density layer, that persists for a long distance after the body.
This behavior was observed before in the framework of micro-scale flow simulations,\cite{mcdonald2011extended}
and can be attributed to the total absence of a heat flux in the 10-moment formulation (see Section~\ref{sec:theory-10-moment-method}).
This discrepancy in the density has a large effect on the wake temperature, that is completely over-predicted (Fig.~\ref{fig:simulations-stagline-Kn0p1}-Top-Centre).
Possible regularizations of the 10-moment systems that can address this issue are discussed in McDonald and Groth, 2008\cite{mcdonald2008extended} and McDonald, 2011.\cite{mcdonald2011extended}

The 14-moment method shows a much higher accuracy.
The density throughout the shock is reproduced accurately, and the stagnation point density is acceptable. 
The temperature field is predicted reasonably, as is the heat flux, except in the immediate vicinity of the flat plate.
Differently than the 10-moment system, the 14-moment system succeeds in reproducing the re-filling of the trail, although some inaccuracy can be appreciated
from the density contours of Fig.~\ref{fig:simulations-Kn0p1-2Dplots}. 
Further comparison of the moment solution to the kinetic solution in the wake region is shown in Appendix~\ref{sec:wake-lateral-profiles}. 

\begin{figure*}[htpb!]
  \centering
  \includegraphics[width=\textwidth]{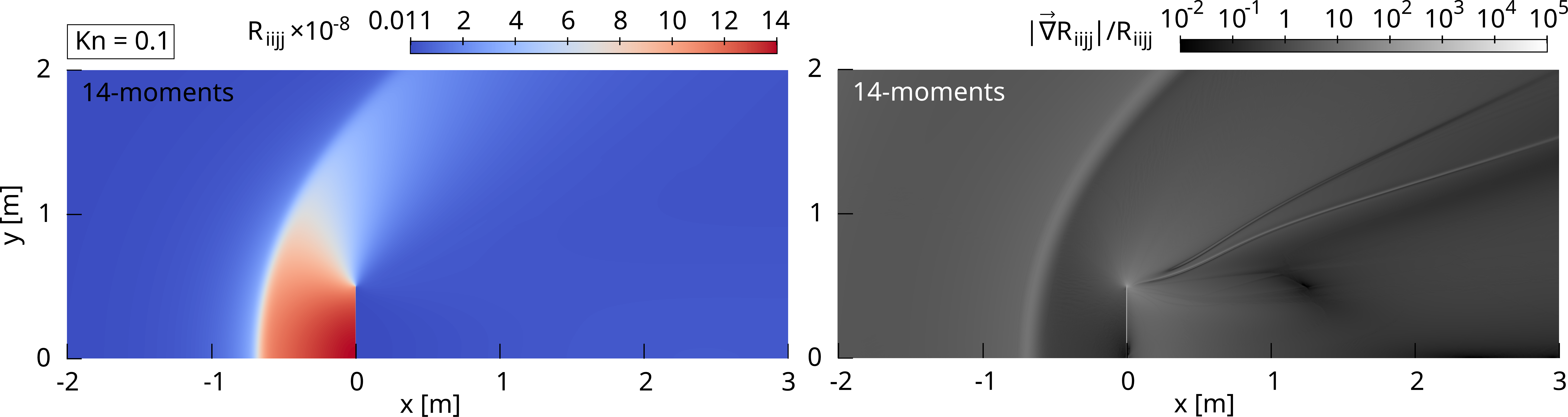}
  \caption{14-moment simulation at  $\mathrm{Kn} = 0.1$. The left box shows the contracted fourth-order moment, $R_{iijj}$, while the right box shows the 
           magnitude of the scaled gradient of $R_{iijj}$, highlighting some low-amplitude discontinuities present in the solution.}
  \label{fig:R_gradR_Kn0p1}
\end{figure*}

When applied to rarefied flow conditions, the 14-moment maximum-entropy method is known to produce a rich set of waves and discontinuities.\cite{boccelli2023rarefied}
These discontinuities do not have a direct kinetic counterpart, and are to be attributed to the mathematical structure of the equations. 
These waves affect predominantly the highest-order moment, $R_{iijj}$, and their effect extends to the lower-order moments as well, although less visibly.
For the considered test case, with $\mathrm{Kn} = 0.1$, these waves do not have an appreciable effect on the lower-order moments, and 
the effect on $R_{iijj}$ itself is small, as shown in Fig.~\ref{fig:R_gradR_Kn0p1}-Left.
However, they can be visualized by computing the normalized spacial gradient of $R_{iijj}$, as done in Fig.~\ref{fig:R_gradR_Kn0p1}-Right.
For the $\mathrm{Kn}=1$ test cases, these discontinuities have a more profound effect, as discussed in Section~\ref{sec:Kn-1-simulations}.

\subsection{Kn = 1 simulations}\label{sec:Kn-1-simulations}

\begin{figure*}[htpb!]
  \centering
  \includegraphics[width=\textwidth]{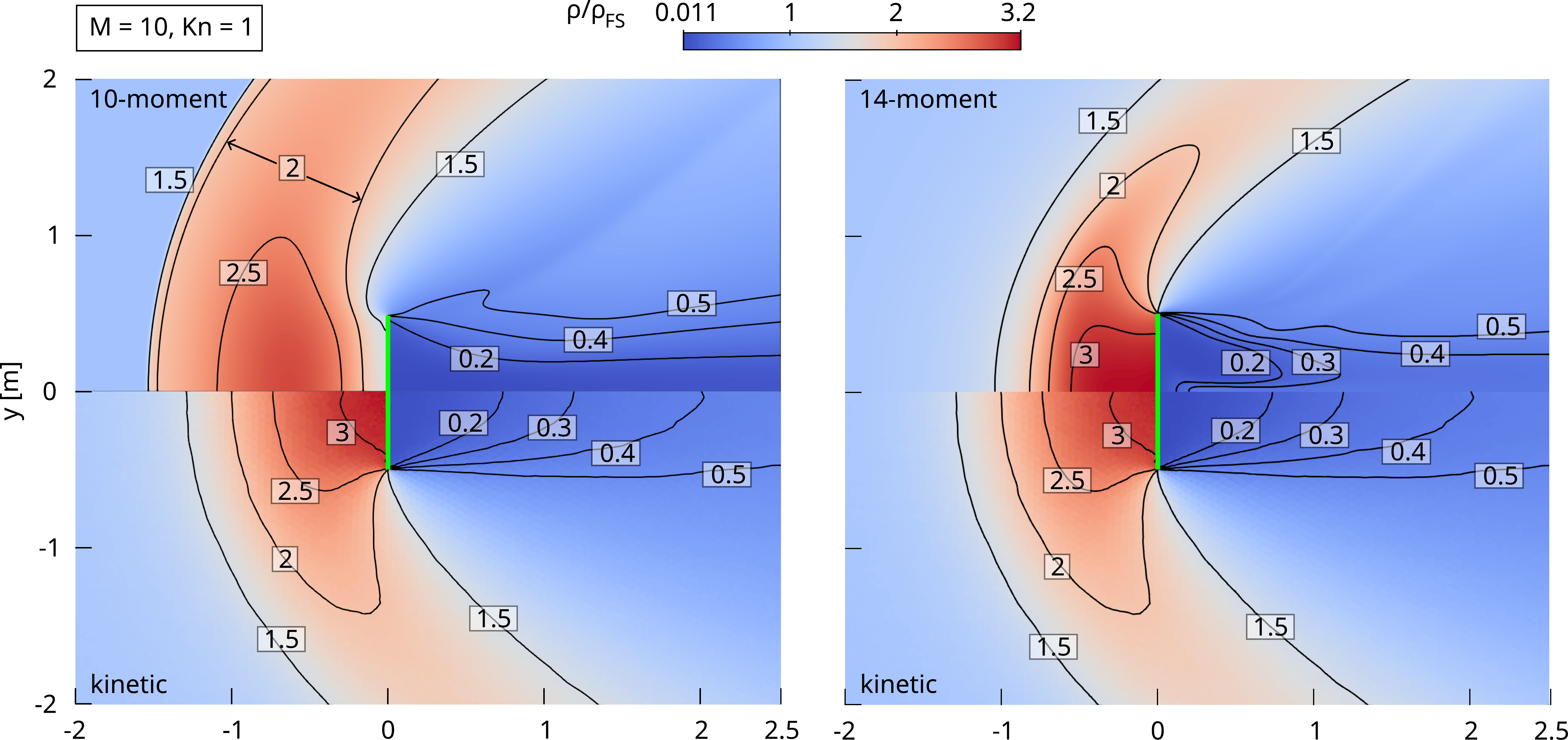}
  \caption{Density field, scaled with the free-stream density. Kinetic simulations (Bottom half of the plots, for $y<0$) compared to the 10-moment (Top-Left) and 14-moment (Top-Right) maximum-entropy simulations, at a Knudsen number Kn = 1.}
  \label{fig:simulations-Kn1-2Dplots}
\end{figure*}

\begin{figure*}[htpb!]
  \centering
  \includegraphics[width=\textwidth]{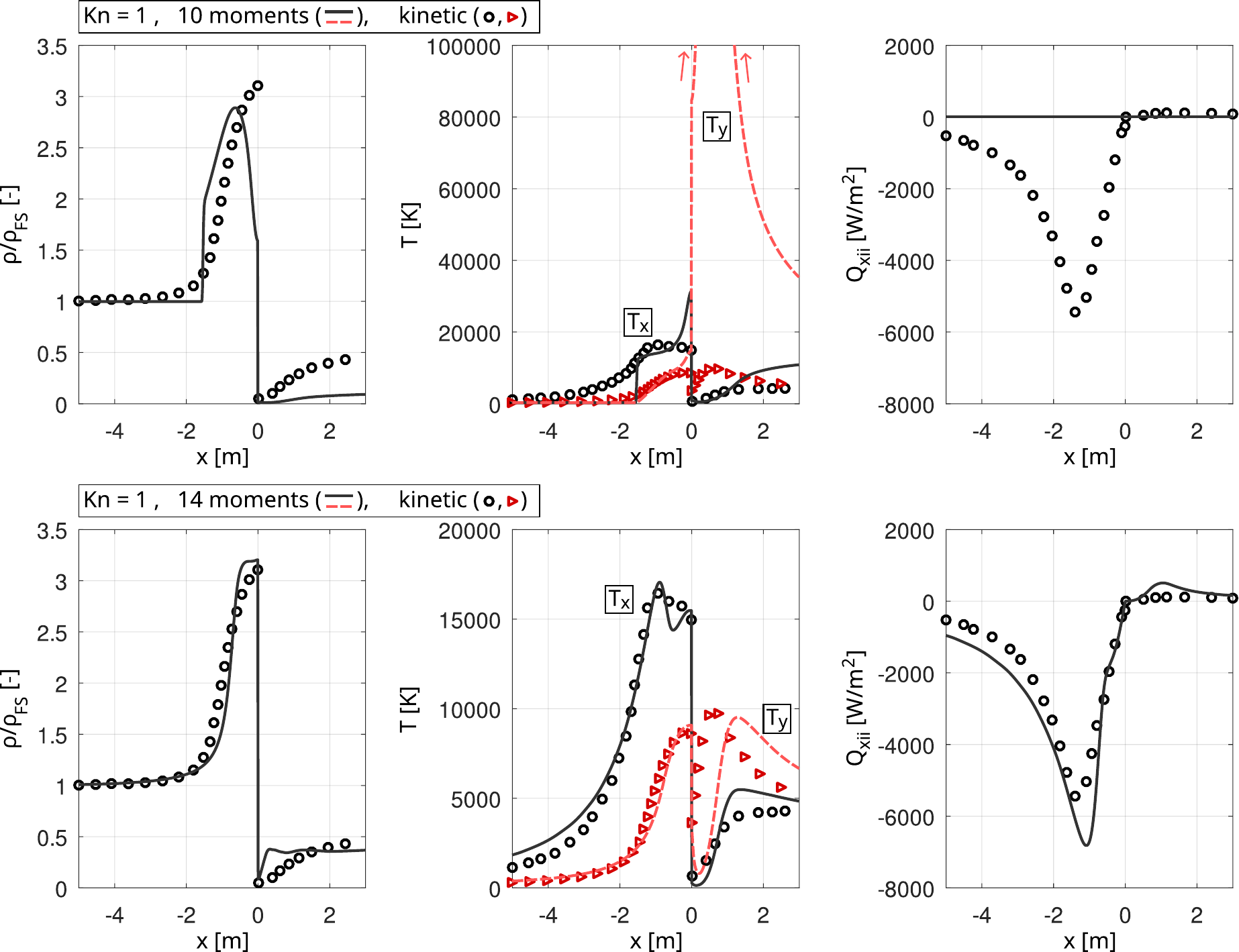}
  \caption{Selected moments along the symmetry line, for the Kn = 1 simulations. 
           Kinetic solution (symbols), 10-moment solution (lines, Top boxes) and 14-moment solution (lines, Bottom boxes).
           Positions with $x < 0$ correspond to the stagnation line, while $x > 0$ denotes the wake.}
  \label{fig:simulations-stagline-Kn1}
\end{figure*}

With respect to the previous simulations, at $\mathrm{Kn}=1$ the kinetic solution shows an extremely diffuse shock, 
to the point that a clear post-shock region is not visible anymore (see Fig.~\ref{fig:simulations-Kn1-2Dplots}).
After crossing the diffuse shock, the flow expands around the flat plate and replenishes the trail, as for the smaller-Knudsen simulations.

In these conditions, the 10 and 14-moment maximum-entropy methods show a markedly reduced accuracy.
The 10-moment method predicts a strong thin shock, followed by a smooth density profile, whose maximum is detached from the body.
Also, the wake does not fill correctly, as already observed in Section~\ref{sec:Kn-0p1-simulations}.

\begin{figure*}[htpb!]
  \centering
  \includegraphics[width=\textwidth]{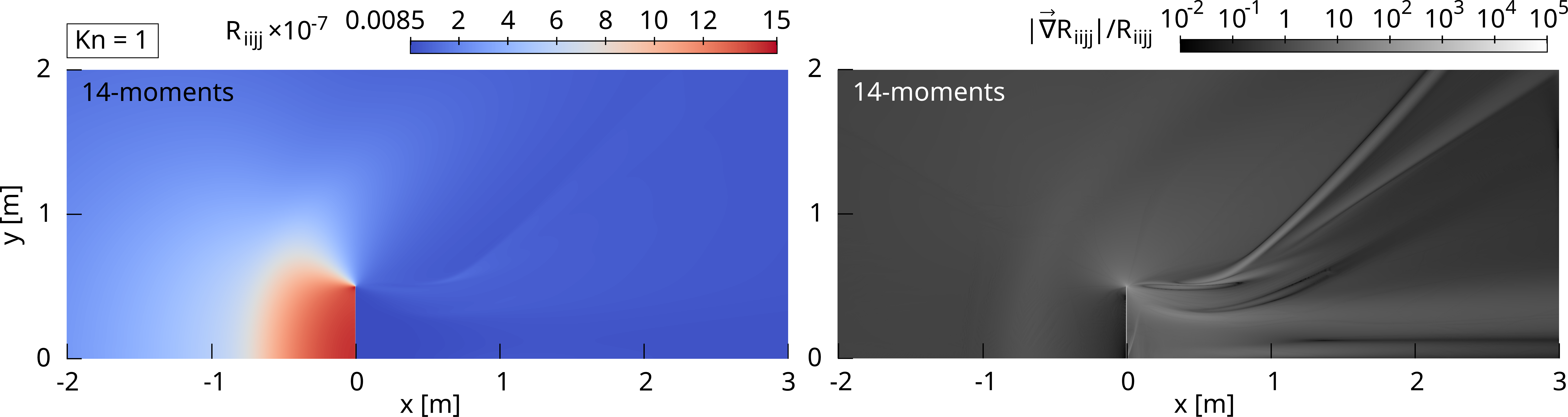}
  \caption{14-moment simulation at  $\mathrm{Kn} = 1$. The left box shows the contracted fourth-order moment, $R_{iijj}$, while the right box shows the 
           magnitude of the scaled gradient of $R_{iijj}$, highlighting some low-amplitude discontinuities present in the solution.}
  \label{fig:R_gradR_Kn1}
\end{figure*}

The 14-moment method, on the other hand, produces a qualitatively reasonable density profile, although with significant deviations from the kinetic solution.
The temperature profiles and the heat flux are predicted also reasonably (see Fig.~\ref{fig:simulations-stagline-Kn1}).
An additional comparison of the profiles, focusing on the wake region, is reported in Appendix~\ref{sec:wake-lateral-profiles}.
The accuracy reduces dramatically in the wake, where a thin higher-density layer creates near the symmetry plane, affecting the density contours.
Providing an intuitive explanation for the presence of this layer, as well as the wiggles that appear in other wake density contour lines (consider for instance 
the contour line at $\rho/\rho_{\mathrm{FS}} = 0.5$, in Fig.~\ref{fig:simulations-Kn1-2Dplots}) is not trivial.
An analysis of the fourth-order moment, $R_{iijj}$, and of its spacial gradient is given in Fig.~\ref{fig:R_gradR_Kn1}, and shows the presence of a rich set of 
waves and discontinuities.
The amplitude of such features is sufficiently large to influence the lower-order moments all the way to the density.

Some of the mentioned waves/discontinuities might be associated with the numerical crossing of the Junk subspace, 
where the flux Jacobian becomes singular. 
In such conditions, the eigenvalues assume a large value and limit the computational time of explicit simulations, via the CFL constraint.
The choice of the limiting parameter, $\sigma_{\mathrm{lim}}$, might play a role in such situations.
However, in the present case, the wiggles in the wake density contours and the anomalously larger-density layer at the symmetry plane, appear to be 
unrelated to this, as the gas state at these locations is observed to be far from the Junk subspace.
Indeed, employing different values of $\sigma_{\mathrm{lim}}$ does not change this topology.
For these reasons, we believe the presented solution to be, not a numerical artifact, but the actual solution of the 14-moment method.
However, additional analysis on this point are suggested as a future work, to further clarify this point.
in particular, we suggest to investigate the effect of (i) different choices of the numerical flux function, (ii) the approximated interpolative closure and 
(iii) the effect of the approximations in the wave speeds.

\section{Distribution functions}\label{sec:VDFs}

\begin{figure*}[htpb!]
  \centering
  \includegraphics[width=\textwidth]{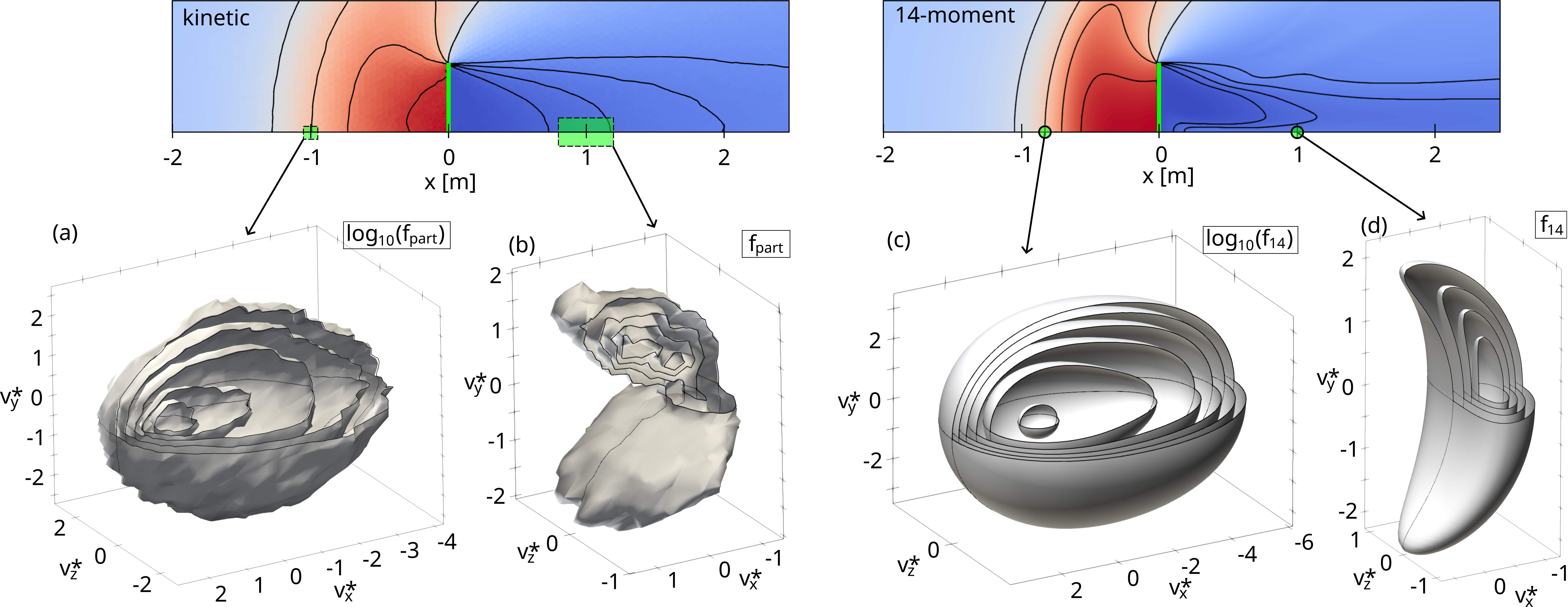}
  \caption{Contours of the velocity distribution function at selected locations in the shock layer and in the wake. 
           Kinetic VDF, reconstructed from particle data (Left) and 14-moment maximum-entropy VDF (Right). 
           Notice that the shock-layer VDFs are represented with logarithmically spaced contours, to aid the visualization, while the wake VDF contours are linearly spaced.}
  \label{fig:VDFs-axis}
\end{figure*}

In this section, we analyze the particle velocity distribution function (VDF) at selected locations inside the shock layer and in the wake.
The VDFs represent the density of particles in velocity space, at a given spacial position.
The VDFs are shown in Fig.~\ref{fig:VDFs-axis}, where the velocity axes are non-dimensionalized as 
$v_i^\star = v_i/\sqrt{p/\rho}$, where the term $\sqrt{p/\rho}$ is proportional to the local thermal velocity.
Also, the analysis is carried in the local frame of reference, where the bulk velocity is zero.
We consider here the particle-based kinetic method and the 14-moment maximum-entropy simulations.
The 10-moment VDFs are not discussed here, because: 
\begin{itemize}
  \item They are ellipsoids and their shape can be easily pictured by considering the ratio of the temperatures, $T_{xx}$ and $T_{yy}$;
  \item The 10-moment system predicts extremely large temperatures in the trail, about 5--10 times larger than the real kinetic value.
        The resulting 10-moment VDFs are thus unphysically spread in velocity space.
\end{itemize}

The VDFs from the kinetic solution are obtained by post-processing the particle data after a steady state is reached, and are shown in Fig.~\ref{fig:VDFs-axis}-Left.
In the shock layer, a sampling region of $(2\times 2)~\si{cm}$, centred around $x=-1~\si{m}$, was sufficient to collect approximately $30\,000$ particles.
In the wake, where the density is considerably lower, a larger sampling region of $(8\times 4)~\si{cm}$, centred around $x = 1~\si{m}$, 
was necessary, and resulted in approximately 8000 particles.

Notice that the sampling regions shown in Fig.~\ref{fig:VDFs-axis}-Left are located across the symmetry line, 
while our simulations formally consider only half of the domain, and exploit symmetry.
Even considering only the particles within the very first cell of the simulated domain would skew the distribution functions towards positive 
velocities, $v_y$, due to the slight offset of the cell from the axis, $y=0$.
This effect is not extreme, but is observable nonetheless.
Therefore, in this section, we artificially introduce into the array of extracted particles, the particles that would theoretically be sampled from the $y<0$ part of the domain.
This is done practically by (i) extracting the simulated particles from the existing numerical half-domain, (ii) doubling them, reversing the $v_y$ velocity component of the twin
particles, and (iii) computing the VDF for this enlarged and symmetrized particle array.
The result is a symmetric VDF, about $v_y$, that is equivalent to running a full-domain simulation.

In the shock layer, the VDF appears to be elongated in the longitudinal direction.
This is a well-known feature of shock waves, where bi-modal distributions are often observed, as a result of the mixing of the cold and fast free-stream particles with 
a slower and warmer post-shock population.\cite{holtz1983molecular,munafo2014spectral}
The wake VDF shows two distinct peaks, at a velocity, $v_y$, that is roughly equal to the thermal velocity ($v_y^\star \approx 1$).
The two peaks are associated with the two streams that expand around the flat plate shoulders, and replenish the wake.
At high-density conditions (small Knudsen number), these streams would interact and result in a recompression shock.
Instead, at the considered low-collisionality, the two streams are able to cross each other at the symmetry plane, interacting only weakly.
For a further discussion of this effect in the framework of high-altitude hypersonic flows, we refer the reader to Bariselli et al, 2018.\cite{bariselli2018aerothermodynamic}

The VDFs associated with the 14-moment method are shown in Fig.~\ref{fig:VDFs-axis}-Right.
As discussed in Section~\ref{sec:theory-14mom-method}, in the present work we do not solve the entropy maximisation directly, during the numerical simulations, as this would 
result in a large computational overhead. 
Instead, we employ a direct approximation of the closing moment.
For this reason, our 14-moment simulations are carried without computing the VDF explicitly, at any point of the simulations.
The 14-moment VDFs shown in this section are obtained a posteriori, by post-processing the simulation results.
For the two selected locations, in the shock layer and in the wake, we extract the gas state from the simulations, and feed it to 
an external optimizer,\cite{boccelli2023gallery} that finds the coefficients, $\bm{\alpha}$, of the 14-moment distribution function (see Eq.~\eqref{eq:f14-equation}).
As the 14-moment and the kinetic simulations do not match exactly, in terms of density field and other moments, we select the probed locations as follows.
In the shock layer, we select the point that has the same density as the probed region of the kinetic solution.
For $\rho/\rho_{\mathrm{FS}}=2$, this point is located at $x \approx -0.8186~\si{m}$.
In the wake, the 14-moment and the kinetic solutions differ sharply, along the axis.
Therefore, for the lack of better criteria, we simply select the same point, $x=1~\si{m}$.

As observed for the kinetic solution, computing the 14-moment VDF from the first cell of the simulated domain would result in a distribution that is (slightly) biased 
towards $v_y$. 
Ideally, one would want to simulate a full domain, and then consider the values of the cell located exactly on the centre.
In this cell, all odd-order moments of $v_y$ (that is, $u_y$, $P_{xy}$, $P_{zy}$ and $Q_{yjj}$) would be zero.
In our half-domain simulations, the first cell centre is located above the centreline, offset by half of a cell size.
Correspondingly, the mentioned moments are slightly non-zero.
For the sake of computing a VDF, we fix here this imbalance by setting these moments to zero, artificially.
Notice that this is equivalent to averaging the cell value with the state of a hypothetical cell located below the symmetry line.

The 14-moment VDFs show analogous features to the particle-based ones.
In the shock layer, the 14-moment VDF is elongated along $v_x$, and is asymmetric, which implies the presence of a heat flux component, $Q_{xjj}$.
As discussed, the wake VDF is expected to contain the contribution of two streams, coming from the two sides of the flat plate.
While the kinetic VDF shows the two different contributions rather clearly, the 14-moment VDF does show the expected curvature, 
but only a single peak---rather than two---is observed.
The problem of low-collisional flows impinging on each other is a common test case for moment methods.
A successful method must be able to reproduce the re-separation of the streams.
The 10-moment method is known to fail this test,\cite{forgues2019higher}
since the associated distribution function cannot assume bi-modal shapes or curvature.
This contributes to the failure of this method in the wake of the flat plate.
On the other hand, the 14-moment method is able to reproduce accurately the re-separation of two jets.\cite{forgues2019higher,boccelli2023rarefied}

It is important to stress that an exact matching between the moment and the kinetic VDFs is not strictly necessary for obtaining accurate predictions of 
the macroscopic quantities---the moments.
However, the opposite is true: 
if the moment method is able to recover the kinetic VDFs accurately, then the macroscopic moment predictions can be expected to be accurate as well.
The discussion is here limited to the $\mathrm{Kn} = 1$ simulations.
From an analysis of the $\mathrm{Kn} = 0.1$ case, the VDFs showed to be qualitatively analogous to the lower-density case, but presenting less non-equilibrium.


\section{Conclusions}\label{sec:conclusions}

As expected, the two investigated maximum-entropy models, the 10 and 14-moment methods, show a different accuracy when applied to high-speed rarefied flows,
depending on the Knudsen number, $\mathrm{Kn}$.
At $\mathrm{Kn}=0.1$, the 10-moment system predicts a reasonable---although inaccurate---shock layer, if compared to a particle-based kinetic solution.
However, the 10-moment simulations predict an extremely elongated wake, composed of a persistent low-density and by an unphysically high temperature, 
due to the lack of a heat flux in the Gaussian formulation.
These issues can be probably mitigated by introducing heat flux closures,\cite{mcdonald2008extended} which is suggested as a future work.
 
The 14-moment simulations show a superior accuracy:
the shock structure is accurately predicted, in terms of shock distance and density profile, 
while the temperature anisotropy and the heat flux are reasonably reproduced. 
The wake profile and the replenishing rate are also reproduced fairly well.
At a larger Knudsen number, $\mathrm{Kn} = 1$, the 10-moment method shows to be inadequate, both in the shock region and in the wake.
On the other hand, the 14-moment method retains a qualitative agreement with the kinetic solution.
Remarkably, the 14-moment method is able to reproduce a fully smooth shock region despite the high Mach number, 
as expected from previous one-dimensional computations.\cite{mcdonald2013affordable}
The largest qualitative discrepancies are observed in the wake, where an unphysical high-density layer is predicted at $\mathrm{Kn=1}$,
sharply affecting the density contours.

Additional insight is given by the analysis of the particle-based and the 14-moment velocity distribution functions.
Inside the shock, as expected, the VDFs are elongated and show a distinctive longitudinal asymmetry, 
suggesting the presence of a strong heat flux.
The particle-based VDFs in the trail show a curved and bi-modal shape, indicating the simultaneous presence of two particle populations,
that are to be identified with the two gas streams that expand over the flat plate shoulders and replenish the trail.
In low-collisional conditions, these streams intersect each other at the centreline, resembling the jet-crossing problem
often studied as a moment method benchmark.\cite{forgues2019higher,desjardins2008quadrature,patel2019three}
The 14-moment VDF mimics this behavior, although in an approximate manner.

Future work suggestions include the formulation of diffusive, instead of specular-reflection, wall boundary conditions, and the 
extension of the presented results to the 21-moment maximum-entropy method, for which an approximated interpolative closure
was recently developed.\cite{giroux2021approximation}


\section*{Data availability statement}
Data sharing is not applicable to this article as no new data were created or analyzed in this study.


\begin{acknowledgments}
The work of SB was funded by the Natural Sciences and Engineering Research Council of Canada (NSERC) through grant number RGPAS-2020-00122.  
The research of PP was funded by a Research Foundation-Flanders (FWO) Strategic Basic PhD fellowship (reference 1S24022N).
This research was supported by the NVIDIA Academic Hardware Acceleration Grant and utilized NVIDIA A100 40GB GPUs.
The authors are very grateful to NSERC, FWO and NVIDIA for the support.
\end{acknowledgments}


\appendix

\section{Approximated closing fluxes for the 14-moment system}

The derivation of the approximated interpolative closing fluxes of the 14-moment system is given in McDonald and Torrilhon, 2013.\cite{mcdonald2013affordable}
Here, we report the final expression for completeness.
First, a parabolic mapping of the moment space is defined, with a parameter
\begin{multline}\label{eq:sigma-mapping-14mom}
  \sigma = \frac{1}{4 P_{ij}P_{ji}} \left\{ 2 P_{ij} P_{ji} + P_{ii}P_{jj} - \rho R_{iijj} \vphantom{\sqrt{P_k^2}} \right. \\
                \left.  + \left[ \left( 2 P_{ij} P_{ji} + P_{ii}P_{jj} - \rho R_{iijj}\right)^2 \right. \right. \\
                    \left. \left. + 8 \rho P_{mn} P_{nm} Q_{kii} \left(P^{-1}\right)_{kl}Q_{ljj}  \right]^{1/2} \right\} \, .
\end{multline}

\noindent The heat flux tensor, $Q_{ijk}$, is approximated as 
\begin{equation}
  Q_{ijk} = K_{ijkm} Q_{mnn} \, ,
\end{equation}

\noindent the fourth-order tensor, $R_{ijkk}$, is written as
\begin{equation}
  R_{ijkk} = \frac{1}{\sigma}Q_{ijl}(P^{-1})_{lm}Q_{mkk} + \frac{2 (1-\sigma) P_{ik} P_{kj} + P_{ij}P_{kk}}{\rho} \, ,
\end{equation}

\noindent and the fifth-order vector, $S_{ijjkk}$, reads 
\begin{multline}
  S_{ijjkk} = \frac{1}{\sigma^2} P_{kn}^{-1}P_{lm}^{-1} Q_{npp} Q_{mjj} Q_{ikl}  \\
+ 2 \sigma^{1/2} \frac{P_{jj} Q_{ikk}}{\rho} + (1 - \sigma^{1/2}) W_{im} Q_{mnn} \, .
\end{multline}

\noindent Notice that $\sigma^{1/2}$ is used here, instead of the value $\sigma^{3/5}$ that appears in the original reference.
A power of $1/2$ was later found to be more accurate. 
The following auxiliary matrices are defined:
\begin{subequations}
\begin{equation}
  B_{lm} = 2 P_{lm}(P^2)_{\alpha \alpha} + 4 (P^3)_{lm} \, ,
\end{equation}
\begin{equation}
  K_{ijkm} = \left[ 2P_{il}(P^2)_{jk} + 2 P_{kl} (P^2)_{ij} + 2 P_{jl}(P^2)_{ik} \right] B_{lm}^{-1} \, ,
\end{equation}
\begin{multline}
  W_{im} = \frac{1}{\rho} \left[ 2 P_{il}(P_{\alpha\alpha})^3 + 12P_{il}(P^3)_{\alpha\alpha} \right. \\ 
      + 14 (P^2)_{\alpha\alpha} (P^2)_{il} + 20 P_{\alpha\alpha}(P^3)_{il} + 20(P^4)_{il} \\ 
\left.- 2(P^2)_{\alpha\alpha}P_{\beta\beta}P_{il} - 6(P_{\alpha\alpha})^2(P^2)_{il}\right]B_{lm}^{-1} \, .
\end{multline}
\end{subequations}

\noindent These equations close the 14-moment system.


\section{Convergence history plots}\label{sec:convergence-history}

\begin{figure}[htpb!]
  \centering
  \includegraphics[width=\columnwidth]{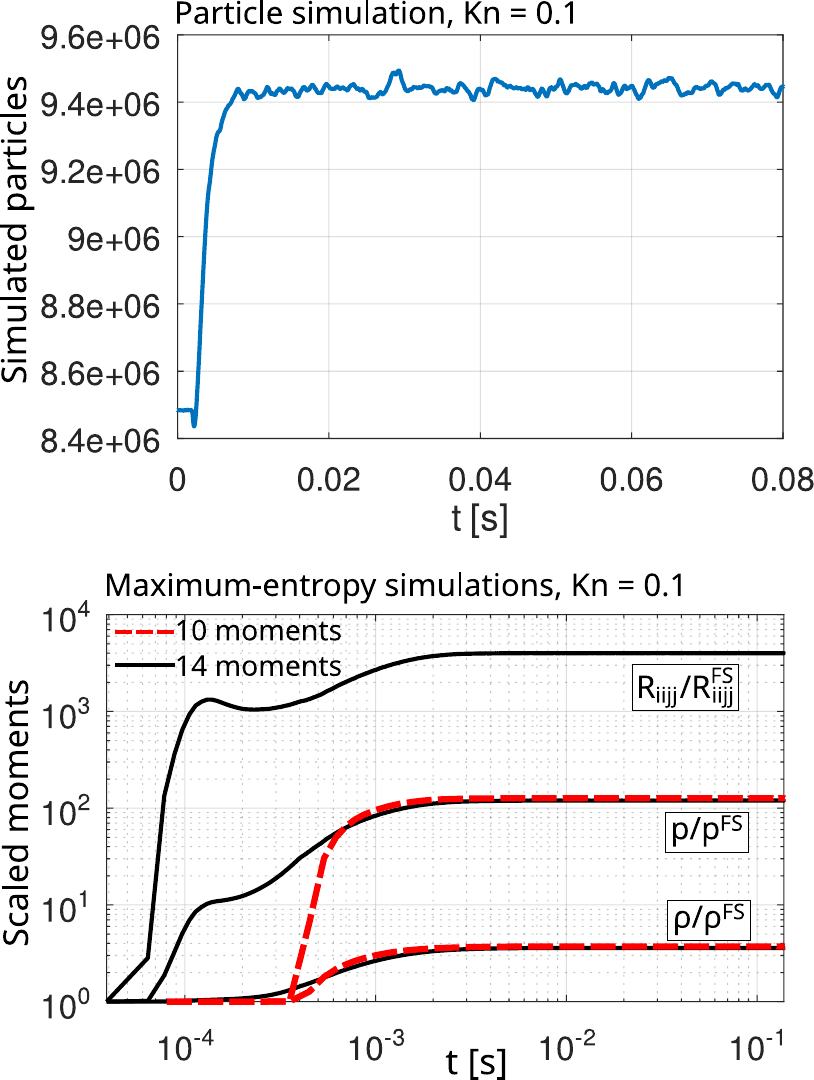}
  \caption{Time convergence plots for the Kn=0.1 case. Top: kinetic simulations (total number of simulated particles).
           Bottom: maximum-entropy simulations (density, pressure and fourth-order moment, scaled with respect to the free-stream values, 
                   at position $x=-0.5~\si{m}$ along the stagnation line).}
  \label{fig:time-convergence-history}
\end{figure}

The particle and maximum-entropy simulations are marched in time until a steady state is reached. 
For the particle simulations, we employed the total number of particles as a criteria for assessing temporal convergence,
as often done in the literature.
Figure~\ref{fig:time-convergence-history}-Top shows the convergence history plot for the $\mathrm{Kn}=0.1$ test case.
After $t = 0.01$--$0.02~\si{s}$ the simulation is converged, but the results showed in this paper are sampled in the range of $t=0.06$--$0.08~\si{s}$
to ensure the convergence of the higher-order moments.
Two-dimensional contour plots of the moments have also been inspected, visually, during the convergence, to ensure the absence 
of unexpected local features.

For maximum-entropy simulations, convergence is assessed both visually and by plotting the time-history of selected points in space.
Figure~\ref{fig:time-convergence-history}-Bottom shows the time convergence of the density, pressure and fourth-order moment, 
at position $x = -0.5~\si{m}$ along the stagnation line.

Spatial convergence is assessed by repeating the simulations on progressively finer grids.

\section{Lateral profiles in the wake}\label{sec:wake-lateral-profiles}

\begin{figure*}[htpb!]
  \centering
  \includegraphics[width=\textwidth]{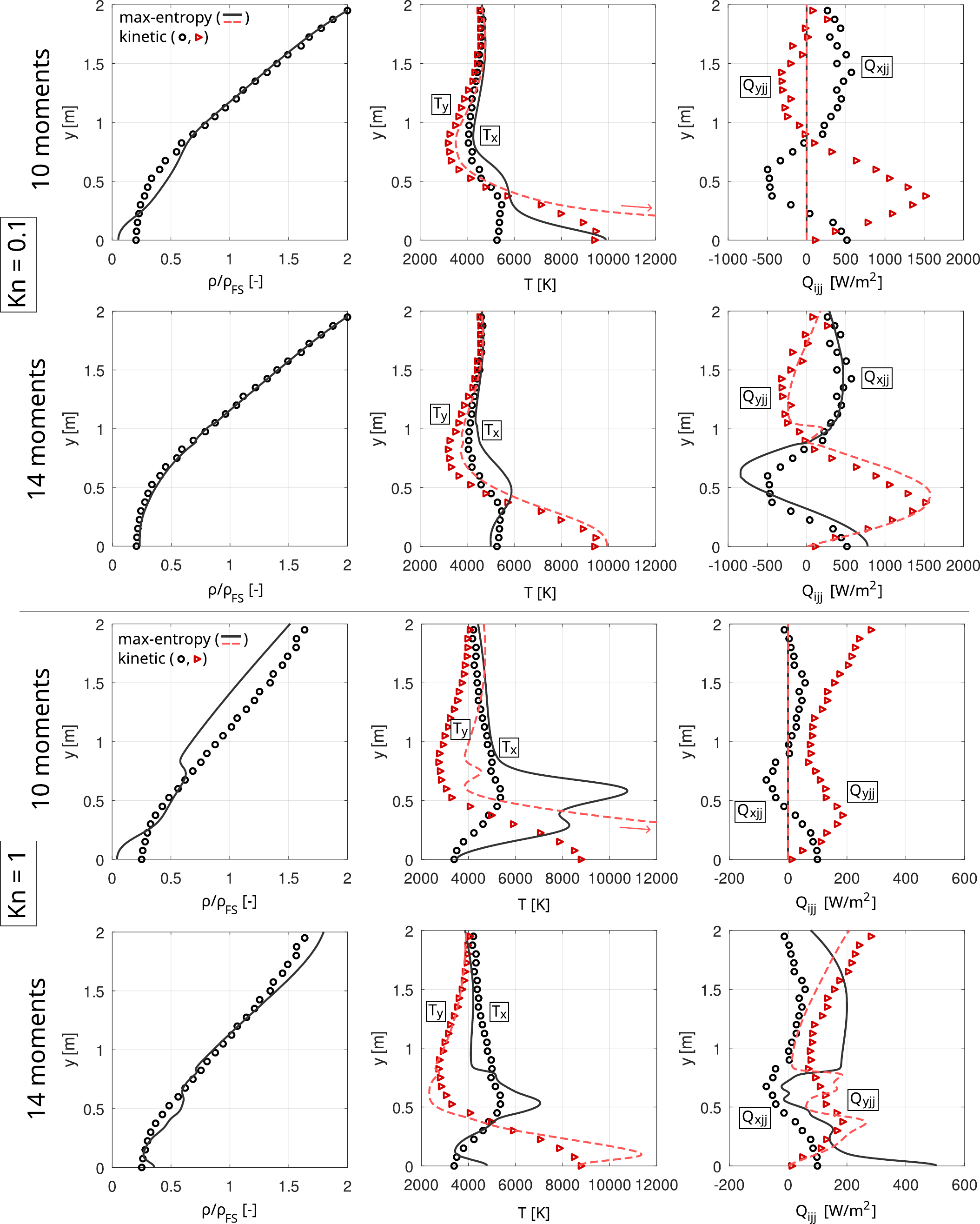}
  \caption{Lateral profiles for selected moments of interest, extracted in the wake region, along the vertical line, $x = 1~\si{m}$. 
           Top plots: 10 and 14-moment cases for $\mathrm{Kn}=0.1$. Bottom: same for $\mathrm{Kn}=1$. 
           The symbols represent the kinetic solution, while the (dashed and solid) lines refer to the maximum-entropy moment solutions.}
  \label{fig:vertical-profiles-wake}
\end{figure*}

Figure~\ref{fig:vertical-profiles-wake} shows the profiles of selected quantities in the wake region, in the $y$-direction. 
the density, $\rho$, normalized with respect to the free-stream density, the temperatures and heat flux components.
The profiles are extracted from the two-dimensional simulations of Fig.~\ref{fig:simulations-Kn0p1-2Dplots} and Fig.~\ref{fig:simulations-Kn1-2Dplots}, along the line $x = 1~\si{m}$.
The accuracy of the moment solutions, with respect to the kinetic solution, is observed to improve for positions located farther in the trail, as collisions
permit a thermalization.
It is interesting to observe how the temperature anisotropy is maximum near the symmetry plane, and decreases as one moves laterally.
Notice that the 10-moment method presents a zero heat flux, by formulation. 
Also notice that the heat flux component, $Q_{yjj}$, predicted by both the kinetic method and the maximum-entropy system, is zero at position $y=0$, 
as expected from symmetry considerations.

\begin{thebibliography}{50}%
\makeatletter
\providecommand \@ifxundefined [1]{%
 \@ifx{#1\undefined}
}%
\providecommand \@ifnum [1]{%
 \ifnum #1\expandafter \@firstoftwo
 \else \expandafter \@secondoftwo
 \fi
}%
\providecommand \@ifx [1]{%
 \ifx #1\expandafter \@firstoftwo
 \else \expandafter \@secondoftwo
 \fi
}%
\providecommand \natexlab [1]{#1}%
\providecommand \enquote  [1]{``#1''}%
\providecommand \bibnamefont  [1]{#1}%
\providecommand \bibfnamefont [1]{#1}%
\providecommand \citenamefont [1]{#1}%
\providecommand \href@noop [0]{\@secondoftwo}%
\providecommand \href [0]{\begingroup \@sanitize@url \@href}%
\providecommand \@href[1]{\@@startlink{#1}\@@href}%
\providecommand \@@href[1]{\endgroup#1\@@endlink}%
\providecommand \@sanitize@url [0]{\catcode `\\12\catcode `\$12\catcode
  `\&12\catcode `\#12\catcode `\^12\catcode `\_12\catcode `\%12\relax}%
\providecommand \@@startlink[1]{}%
\providecommand \@@endlink[0]{}%
\providecommand \url  [0]{\begingroup\@sanitize@url \@url }%
\providecommand \@url [1]{\endgroup\@href {#1}{\urlprefix }}%
\providecommand \urlprefix  [0]{URL }%
\providecommand \Eprint [0]{\href }%
\providecommand \doibase [0]{https://doi.org/}%
\providecommand \selectlanguage [0]{\@gobble}%
\providecommand \bibinfo  [0]{\@secondoftwo}%
\providecommand \bibfield  [0]{\@secondoftwo}%
\providecommand \translation [1]{[#1]}%
\providecommand \BibitemOpen [0]{}%
\providecommand \bibitemStop [0]{}%
\providecommand \bibitemNoStop [0]{.\EOS\space}%
\providecommand \EOS [0]{\spacefactor3000\relax}%
\providecommand \BibitemShut  [1]{\csname bibitem#1\endcsname}%
\let\auto@bib@innerbib\@empty
\bibitem [{\citenamefont {Lofthouse}, \citenamefont {Boyd},\ and\ \citenamefont
  {Wright}(2007)}]{lofthouse2007effects}%
  \BibitemOpen
  \bibfield  {author} {\bibinfo {author} {\bibfnamefont {A.~J.}\ \bibnamefont
  {Lofthouse}}, \bibinfo {author} {\bibfnamefont {I.~D.}\ \bibnamefont
  {Boyd}},\ and\ \bibinfo {author} {\bibfnamefont {M.~J.}\ \bibnamefont
  {Wright}},\ }\bibfield  {title} {\enquote {\bibinfo {title} {Effects of
  continuum breakdown on hypersonic aerothermodynamics},}\ }\href@noop {}
  {\bibfield  {journal} {\bibinfo  {journal} {Physics of Fluids}\ }\textbf
  {\bibinfo {volume} {19}},\ \bibinfo {pages} {027105} (\bibinfo {year}
  {2007})}\BibitemShut {NoStop}%
\bibitem [{\citenamefont {Rebrov}(2001)}]{rebrov2001free}%
  \BibitemOpen
  \bibfield  {author} {\bibinfo {author} {\bibfnamefont {A.}~\bibnamefont
  {Rebrov}},\ }\bibfield  {title} {\enquote {\bibinfo {title} {Free jets in
  vacuum technologies},}\ }\href@noop {} {\bibfield  {journal} {\bibinfo
  {journal} {Journal of Vacuum Science \& Technology A: Vacuum, Surfaces, and
  Films}\ }\textbf {\bibinfo {volume} {19}},\ \bibinfo {pages} {1679--1687}
  (\bibinfo {year} {2001})}\BibitemShut {NoStop}%
\bibitem [{\citenamefont {Ferziger}\ and\ \citenamefont
  {Kaper}(1972)}]{ferziger1972mathematical}%
  \BibitemOpen
  \bibfield  {author} {\bibinfo {author} {\bibfnamefont {J.~H.}\ \bibnamefont
  {Ferziger}}\ and\ \bibinfo {author} {\bibfnamefont {H.~G.}\ \bibnamefont
  {Kaper}},\ }\href@noop {} {\emph {\bibinfo {title} {Mathematical theory of
  transport processes in gases}}}\ (\bibinfo  {publisher} {North-Holland
  Publishing Company},\ \bibinfo {year} {1972})\BibitemShut {NoStop}%
\bibitem [{\citenamefont {Josyula}\ and\ \citenamefont
  {Burt}(2011)}]{josyula2011review}%
  \BibitemOpen
  \bibfield  {author} {\bibinfo {author} {\bibfnamefont {E.}~\bibnamefont
  {Josyula}}\ and\ \bibinfo {author} {\bibfnamefont {J.}~\bibnamefont {Burt}},\
  }\href@noop {} {\enquote {\bibinfo {title} {Review of rarefied gas effects in
  hypersonic applications},}\ }\bibinfo {type} {Tech. Rep.}\ (\bibinfo
  {institution} {AIR FORCE RESEARCH LAB WRIGHT-PATTERSON AFB OH},\ \bibinfo
  {year} {2011})\BibitemShut {NoStop}%
\bibitem [{\citenamefont {Bird}(1994)}]{bird1994molecular}%
  \BibitemOpen
  \bibfield  {author} {\bibinfo {author} {\bibfnamefont {G.~A.}\ \bibnamefont
  {Bird}},\ }\href@noop {} {\emph {\bibinfo {title} {Molecular gas dynamics and
  the direct simulation of gas flows}}}\ (\bibinfo  {publisher} {Clarendon
  Press},\ \bibinfo {year} {1994})\BibitemShut {NoStop}%
\bibitem [{\citenamefont {Mieussens}(2000)}]{mieussens2000discrete}%
  \BibitemOpen
  \bibfield  {author} {\bibinfo {author} {\bibfnamefont {L.}~\bibnamefont
  {Mieussens}},\ }\bibfield  {title} {\enquote {\bibinfo {title}
  {Discrete-velocity models and numerical schemes for the boltzmann-bgk
  equation in plane and axisymmetric geometries},}\ }\href@noop {} {\bibfield
  {journal} {\bibinfo  {journal} {Journal of Computational Physics}\ }\textbf
  {\bibinfo {volume} {162}},\ \bibinfo {pages} {429--466} (\bibinfo {year}
  {2000})}\BibitemShut {NoStop}%
\bibitem [{\citenamefont {Aristov}, \citenamefont {Voronich},\ and\
  \citenamefont {Zabelok}(2019)}]{aristov2019direct}%
  \BibitemOpen
  \bibfield  {author} {\bibinfo {author} {\bibfnamefont {V.}~\bibnamefont
  {Aristov}}, \bibinfo {author} {\bibfnamefont {I.}~\bibnamefont {Voronich}},\
  and\ \bibinfo {author} {\bibfnamefont {S.}~\bibnamefont {Zabelok}},\
  }\bibfield  {title} {\enquote {\bibinfo {title} {Direct methods for solving
  the boltzmann equations: Comparisons with direct simulation monte carlo and
  possibilities},}\ }\href@noop {} {\bibfield  {journal} {\bibinfo  {journal}
  {Physics of Fluids}\ }\textbf {\bibinfo {volume} {31}} (\bibinfo {year}
  {2019})}\BibitemShut {NoStop}%
\bibitem [{\citenamefont {Kim}, \citenamefont {Gorji},\ and\ \citenamefont
  {Jun}(2023)}]{kim2023critical}%
  \BibitemOpen
  \bibfield  {author} {\bibinfo {author} {\bibfnamefont {S.}~\bibnamefont
  {Kim}}, \bibinfo {author} {\bibfnamefont {H.}~\bibnamefont {Gorji}},\ and\
  \bibinfo {author} {\bibfnamefont {E.}~\bibnamefont {Jun}},\ }\bibfield
  {title} {\enquote {\bibinfo {title} {Critical assessment of various particle
  fokker--planck models for monatomic rarefied gas flows},}\ }\href@noop {}
  {\bibfield  {journal} {\bibinfo  {journal} {Physics of Fluids}\ }\textbf
  {\bibinfo {volume} {35}} (\bibinfo {year} {2023})}\BibitemShut {NoStop}%
\bibitem [{\citenamefont {Frezzotti}, \citenamefont {Ghiroldi},\ and\
  \citenamefont {Gibelli}(2011)}]{frezzotti2011solving}%
  \BibitemOpen
  \bibfield  {author} {\bibinfo {author} {\bibfnamefont {A.}~\bibnamefont
  {Frezzotti}}, \bibinfo {author} {\bibfnamefont {G.~P.}\ \bibnamefont
  {Ghiroldi}},\ and\ \bibinfo {author} {\bibfnamefont {L.}~\bibnamefont
  {Gibelli}},\ }\bibfield  {title} {\enquote {\bibinfo {title} {Solving the
  boltzmann equation on gpus},}\ }\href@noop {} {\bibfield  {journal} {\bibinfo
   {journal} {Computer Physics Communications}\ }\textbf {\bibinfo {volume}
  {182}},\ \bibinfo {pages} {2445--2453} (\bibinfo {year} {2011})}\BibitemShut
  {NoStop}%
\bibitem [{\citenamefont {Torrilhon}(2016)}]{torrilhon2016modeling}%
  \BibitemOpen
  \bibfield  {author} {\bibinfo {author} {\bibfnamefont {M.}~\bibnamefont
  {Torrilhon}},\ }\bibfield  {title} {\enquote {\bibinfo {title} {Modeling
  nonequilibrium gas flow based on moment equations},}\ }\href@noop {}
  {\bibfield  {journal} {\bibinfo  {journal} {Annual review of fluid
  mechanics}\ }\textbf {\bibinfo {volume} {48}},\ \bibinfo {pages} {429--458}
  (\bibinfo {year} {2016})}\BibitemShut {NoStop}%
\bibitem [{\citenamefont {Grad}(1949)}]{grad1949kinetic}%
  \BibitemOpen
  \bibfield  {author} {\bibinfo {author} {\bibfnamefont {H.}~\bibnamefont
  {Grad}},\ }\bibfield  {title} {\enquote {\bibinfo {title} {On the kinetic
  theory of rarefied gases},}\ }\href@noop {} {\bibfield  {journal} {\bibinfo
  {journal} {Communications on pure and applied mathematics}\ }\textbf
  {\bibinfo {volume} {2}},\ \bibinfo {pages} {331--407} (\bibinfo {year}
  {1949})}\BibitemShut {NoStop}%
\bibitem [{\citenamefont {Struchtrup}\ and\ \citenamefont
  {Torrilhon}(2003)}]{struchtrup2003regularization}%
  \BibitemOpen
  \bibfield  {author} {\bibinfo {author} {\bibfnamefont {H.}~\bibnamefont
  {Struchtrup}}\ and\ \bibinfo {author} {\bibfnamefont {M.}~\bibnamefont
  {Torrilhon}},\ }\bibfield  {title} {\enquote {\bibinfo {title}
  {Regularization of {Grad}’s 13 moment equations: Derivation and linear
  analysis},}\ }\href@noop {} {\bibfield  {journal} {\bibinfo  {journal}
  {Physics of Fluids}\ }\textbf {\bibinfo {volume} {15}},\ \bibinfo {pages}
  {2668--2680} (\bibinfo {year} {2003})}\BibitemShut {NoStop}%
\bibitem [{\citenamefont {Cai}, \citenamefont {Fan},\ and\ \citenamefont
  {Li}(2014)}]{cai2014globally}%
  \BibitemOpen
  \bibfield  {author} {\bibinfo {author} {\bibfnamefont {Z.}~\bibnamefont
  {Cai}}, \bibinfo {author} {\bibfnamefont {Y.}~\bibnamefont {Fan}},\ and\
  \bibinfo {author} {\bibfnamefont {R.}~\bibnamefont {Li}},\ }\bibfield
  {title} {\enquote {\bibinfo {title} {Globally hyperbolic regularization of
  {Grad}'s moment system},}\ }\href@noop {} {\bibfield  {journal} {\bibinfo
  {journal} {Communications on pure and applied mathematics}\ }\textbf
  {\bibinfo {volume} {67}},\ \bibinfo {pages} {464--518} (\bibinfo {year}
  {2014})}\BibitemShut {NoStop}%
\bibitem [{\citenamefont {Fox}(2009)}]{fox2009higher}%
  \BibitemOpen
  \bibfield  {author} {\bibinfo {author} {\bibfnamefont {R.~O.}\ \bibnamefont
  {Fox}},\ }\bibfield  {title} {\enquote {\bibinfo {title} {Higher-order
  quadrature-based moment methods for kinetic equations},}\ }\href@noop {}
  {\bibfield  {journal} {\bibinfo  {journal} {Journal of Computational
  Physics}\ }\textbf {\bibinfo {volume} {228}},\ \bibinfo {pages} {7771--7791}
  (\bibinfo {year} {2009})}\BibitemShut {NoStop}%
\bibitem [{\citenamefont {Struchtrup}(2005)}]{struchtrup2005macroscopic}%
  \BibitemOpen
  \bibfield  {author} {\bibinfo {author} {\bibfnamefont {H.}~\bibnamefont
  {Struchtrup}},\ }\href@noop {} {\emph {\bibinfo {title} {Macroscopic
  transport equations for rarefied gas flows: approximation methods in kinetic
  theory}}}\ (\bibinfo  {publisher} {Springer Science \& Business Media},\
  \bibinfo {year} {2005})\BibitemShut {NoStop}%
\bibitem [{\citenamefont {M{\"u}ller}\ and\ \citenamefont
  {Ruggeri}(1993)}]{muller1993extended}%
  \BibitemOpen
  \bibfield  {author} {\bibinfo {author} {\bibfnamefont {I.}~\bibnamefont
  {M{\"u}ller}}\ and\ \bibinfo {author} {\bibfnamefont {T.}~\bibnamefont
  {Ruggeri}},\ }\href@noop {} {\emph {\bibinfo {title} {Extended
  thermodynamics}}}\ (\bibinfo  {publisher} {Springer-Verlag},\ \bibinfo {year}
  {1993})\BibitemShut {NoStop}%
\bibitem [{\citenamefont {Levermore}(1996)}]{levermore1996moment}%
  \BibitemOpen
  \bibfield  {author} {\bibinfo {author} {\bibfnamefont {C.~D.}\ \bibnamefont
  {Levermore}},\ }\bibfield  {title} {\enquote {\bibinfo {title} {Moment
  closure hierarchies for kinetic theories},}\ }\href@noop {} {\bibfield
  {journal} {\bibinfo  {journal} {Journal of statistical Physics}\ }\textbf
  {\bibinfo {volume} {83}},\ \bibinfo {pages} {1021--1065} (\bibinfo {year}
  {1996})}\BibitemShut {NoStop}%
\bibitem [{\citenamefont {Forgues}\ \emph {et~al.}(2019)\citenamefont
  {Forgues}, \citenamefont {Ivan}, \citenamefont {Trottier},\ and\
  \citenamefont {McDonald}}]{forgues2019gaussian}%
  \BibitemOpen
  \bibfield  {author} {\bibinfo {author} {\bibfnamefont {F.}~\bibnamefont
  {Forgues}}, \bibinfo {author} {\bibfnamefont {L.}~\bibnamefont {Ivan}},
  \bibinfo {author} {\bibfnamefont {A.}~\bibnamefont {Trottier}},\ and\
  \bibinfo {author} {\bibfnamefont {J.~G.}\ \bibnamefont {McDonald}},\
  }\bibfield  {title} {\enquote {\bibinfo {title} {A {Gaussian} moment method
  for polydisperse multiphase flow modelling},}\ }\href@noop {} {\bibfield
  {journal} {\bibinfo  {journal} {Journal of Computational Physics}\ }\textbf
  {\bibinfo {volume} {398}},\ \bibinfo {pages} {108839} (\bibinfo {year}
  {2019})}\BibitemShut {NoStop}%
\bibitem [{\citenamefont {Boccelli}\ \emph {et~al.}(2023)\citenamefont
  {Boccelli}, \citenamefont {Kaufmann}, \citenamefont {Magin},\ and\
  \citenamefont {McDonald}}]{boccelli2023rarefied}%
  \BibitemOpen
  \bibfield  {author} {\bibinfo {author} {\bibfnamefont {S.}~\bibnamefont
  {Boccelli}}, \bibinfo {author} {\bibfnamefont {W.}~\bibnamefont {Kaufmann}},
  \bibinfo {author} {\bibfnamefont {T.~E.}\ \bibnamefont {Magin}},\ and\
  \bibinfo {author} {\bibfnamefont {J.~G.}\ \bibnamefont {McDonald}},\
  }\bibfield  {title} {\enquote {\bibinfo {title} {Numerical simulation of
  rarefied supersonic flows using a fourth-order maximum-entropy moment method
  with interpolative closure},}\ }\href@noop {} {\bibfield  {journal} {\bibinfo
   {journal} {Under review}\ } (\bibinfo {year} {2023})}\BibitemShut {NoStop}%
\bibitem [{\citenamefont {Jayaraman}, \citenamefont {Liu},\ and\ \citenamefont
  {Panesi}(2017)}]{jayaraman2017multi}%
  \BibitemOpen
  \bibfield  {author} {\bibinfo {author} {\bibfnamefont {V.}~\bibnamefont
  {Jayaraman}}, \bibinfo {author} {\bibfnamefont {Y.}~\bibnamefont {Liu}},\
  and\ \bibinfo {author} {\bibfnamefont {M.}~\bibnamefont {Panesi}},\
  }\bibfield  {title} {\enquote {\bibinfo {title} {Multi-group maximum entropy
  model for translational non-equilibrium},}\ }in\ \href@noop {} {\emph
  {\bibinfo {booktitle} {47th AIAA Thermophysics Conference}}}\ (\bibinfo
  {year} {2017})\ p.\ \bibinfo {pages} {4024}\BibitemShut {NoStop}%
\bibitem [{\citenamefont {Romano}(2002)}]{romano20022d}%
  \BibitemOpen
  \bibfield  {author} {\bibinfo {author} {\bibfnamefont {V.}~\bibnamefont
  {Romano}},\ }\bibfield  {title} {\enquote {\bibinfo {title} {{2D} simulation
  of a silicon {MESFET} with a nonparabolic hydrodynamical model based on the
  maximum entropy principle},}\ }\href@noop {} {\bibfield  {journal} {\bibinfo
  {journal} {Journal of Computational Physics}\ }\textbf {\bibinfo {volume}
  {176}},\ \bibinfo {pages} {70--92} (\bibinfo {year} {2002})}\BibitemShut
  {NoStop}%
\bibitem [{\citenamefont {Bhatnagar}, \citenamefont {Gross},\ and\
  \citenamefont {Krook}(1954)}]{bhatnagar1954model}%
  \BibitemOpen
  \bibfield  {author} {\bibinfo {author} {\bibfnamefont {P.~L.}\ \bibnamefont
  {Bhatnagar}}, \bibinfo {author} {\bibfnamefont {E.~P.}\ \bibnamefont
  {Gross}},\ and\ \bibinfo {author} {\bibfnamefont {M.}~\bibnamefont {Krook}},\
  }\bibfield  {title} {\enquote {\bibinfo {title} {A model for collision
  processes in gases. {I}. small amplitude processes in charged and neutral
  one-component systems},}\ }\href@noop {} {\bibfield  {journal} {\bibinfo
  {journal} {Physical review}\ }\textbf {\bibinfo {volume} {94}},\ \bibinfo
  {pages} {511} (\bibinfo {year} {1954})}\BibitemShut {NoStop}%
\bibitem [{\citenamefont {Gallis}\ and\ \citenamefont
  {Torczynski}(2000)}]{gallis2000application}%
  \BibitemOpen
  \bibfield  {author} {\bibinfo {author} {\bibfnamefont {M.}~\bibnamefont
  {Gallis}}\ and\ \bibinfo {author} {\bibfnamefont {J.}~\bibnamefont
  {Torczynski}},\ }\bibfield  {title} {\enquote {\bibinfo {title} {The
  application of the bgk model in particle simulations},}\ }in\ \href@noop {}
  {\emph {\bibinfo {booktitle} {34th Thermophysics Conference}}}\ (\bibinfo
  {year} {2000})\ p.\ \bibinfo {pages} {2360}\BibitemShut {NoStop}%
\bibitem [{\citenamefont {Macrossan}(2001)}]{macrossan2001particle}%
  \BibitemOpen
  \bibfield  {author} {\bibinfo {author} {\bibfnamefont {M.}~\bibnamefont
  {Macrossan}},\ }\bibfield  {title} {\enquote {\bibinfo {title} {A particle
  simulation method for the bgk equation},}\ }in\ \href@noop {} {\emph
  {\bibinfo {booktitle} {AIP Conference Proceedings}}},\ Vol.\ \bibinfo
  {volume} {585}\ (\bibinfo {organization} {American Institute of Physics},\
  \bibinfo {year} {2001})\ pp.\ \bibinfo {pages} {426--433}\BibitemShut
  {NoStop}%
\bibitem [{\citenamefont {Dreyer}(1987)}]{dreyer1987maximisation}%
  \BibitemOpen
  \bibfield  {author} {\bibinfo {author} {\bibfnamefont {W.}~\bibnamefont
  {Dreyer}},\ }\bibfield  {title} {\enquote {\bibinfo {title} {Maximisation of
  the entropy in non-equilibrium},}\ }\href@noop {} {\bibfield  {journal}
  {\bibinfo  {journal} {Journal of Physics A: Mathematical and General}\
  }\textbf {\bibinfo {volume} {20}},\ \bibinfo {pages} {6505} (\bibinfo {year}
  {1987})}\BibitemShut {NoStop}%
\bibitem [{\citenamefont {McDonald}\ and\ \citenamefont
  {Groth}(2005)}]{mcdonald2005numerical}%
  \BibitemOpen
  \bibfield  {author} {\bibinfo {author} {\bibfnamefont {J.}~\bibnamefont
  {McDonald}}\ and\ \bibinfo {author} {\bibfnamefont {C.}~\bibnamefont
  {Groth}},\ }\bibfield  {title} {\enquote {\bibinfo {title} {Numerical
  modeling of micron-scale flows using the {Gaussian} moment closure},}\ }in\
  \href@noop {} {\emph {\bibinfo {booktitle} {35th AIAA Fluid Dynamics
  Conference and Exhibit}}}\ (\bibinfo {year} {2005})\ p.\ \bibinfo {pages}
  {5035}\BibitemShut {NoStop}%
\bibitem [{\citenamefont {Boccelli}\ \emph {et~al.}(2020)\citenamefont
  {Boccelli}, \citenamefont {Giroux}, \citenamefont {Magin}, \citenamefont
  {Groth},\ and\ \citenamefont {McDonald}}]{boccelli202014}%
  \BibitemOpen
  \bibfield  {author} {\bibinfo {author} {\bibfnamefont {S.}~\bibnamefont
  {Boccelli}}, \bibinfo {author} {\bibfnamefont {F.}~\bibnamefont {Giroux}},
  \bibinfo {author} {\bibfnamefont {T.~E.}\ \bibnamefont {Magin}}, \bibinfo
  {author} {\bibfnamefont {C.}~\bibnamefont {Groth}},\ and\ \bibinfo {author}
  {\bibfnamefont {J.~G.}\ \bibnamefont {McDonald}},\ }\bibfield  {title}
  {\enquote {\bibinfo {title} {A 14-moment maximum-entropy description of
  electrons in crossed electric and magnetic fields},}\ }\href@noop {}
  {\bibfield  {journal} {\bibinfo  {journal} {Physics of Plasmas}\ }\textbf
  {\bibinfo {volume} {27}},\ \bibinfo {pages} {123506} (\bibinfo {year}
  {2020})}\BibitemShut {NoStop}%
\bibitem [{\citenamefont {Boccelli}, \citenamefont {Giroux},\ and\
  \citenamefont {McDonald}(2023)}]{boccelli2023gallery}%
  \BibitemOpen
  \bibfield  {author} {\bibinfo {author} {\bibfnamefont {S.}~\bibnamefont
  {Boccelli}}, \bibinfo {author} {\bibfnamefont {F.}~\bibnamefont {Giroux}},\
  and\ \bibinfo {author} {\bibfnamefont {J.~G.}\ \bibnamefont {McDonald}},\
  }\bibfield  {title} {\enquote {\bibinfo {title} {A gallery of maximum-entropy
  distributions: 14 and 21 moments},}\ }\href@noop {} {\bibfield  {journal}
  {\bibinfo  {journal} {In preparation}\ } (\bibinfo {year}
  {2023})}\BibitemShut {NoStop}%
\bibitem [{\citenamefont {McDonald}\ and\ \citenamefont
  {Torrilhon}(2013)}]{mcdonald2013affordable}%
  \BibitemOpen
  \bibfield  {author} {\bibinfo {author} {\bibfnamefont {J.}~\bibnamefont
  {McDonald}}\ and\ \bibinfo {author} {\bibfnamefont {M.}~\bibnamefont
  {Torrilhon}},\ }\bibfield  {title} {\enquote {\bibinfo {title} {Affordable
  robust moment closures for {CFD} based on the maximum-entropy hierarchy},}\
  }\href@noop {} {\bibfield  {journal} {\bibinfo  {journal} {Journal of
  Computational Physics}\ }\textbf {\bibinfo {volume} {251}},\ \bibinfo {pages}
  {500--523} (\bibinfo {year} {2013})}\BibitemShut {NoStop}%
\bibitem [{\citenamefont {Schaerer}, \citenamefont {Bansal},\ and\
  \citenamefont {Torrilhon}(2017)}]{schaerer2017efficient}%
  \BibitemOpen
  \bibfield  {author} {\bibinfo {author} {\bibfnamefont {R.~P.}\ \bibnamefont
  {Schaerer}}, \bibinfo {author} {\bibfnamefont {P.}~\bibnamefont {Bansal}},\
  and\ \bibinfo {author} {\bibfnamefont {M.}~\bibnamefont {Torrilhon}},\
  }\bibfield  {title} {\enquote {\bibinfo {title} {Efficient algorithms and
  implementations of entropy-based moment closures for rarefied gases},}\
  }\href@noop {} {\bibfield  {journal} {\bibinfo  {journal} {Journal of
  Computational Physics}\ }\textbf {\bibinfo {volume} {340}},\ \bibinfo {pages}
  {138--159} (\bibinfo {year} {2017})}\BibitemShut {NoStop}%
\bibitem [{\citenamefont {Zheng}, \citenamefont {Yang},\ and\ \citenamefont
  {Chen}(2023)}]{zheng2023stabilizing}%
  \BibitemOpen
  \bibfield  {author} {\bibinfo {author} {\bibfnamefont {C.}~\bibnamefont
  {Zheng}}, \bibinfo {author} {\bibfnamefont {W.}~\bibnamefont {Yang}},\ and\
  \bibinfo {author} {\bibfnamefont {S.}~\bibnamefont {Chen}},\ }\bibfield
  {title} {\enquote {\bibinfo {title} {Stabilizing the maximal entropy moment
  method for rarefied gas dynamics at single-precision},}\ }\href@noop {}
  {\bibfield  {journal} {\bibinfo  {journal} {arXiv preprint arXiv:2303.02898}\
  } (\bibinfo {year} {2023})}\BibitemShut {NoStop}%
\bibitem [{Note1()}]{Note1}%
  \BibitemOpen
  \bibinfo {note} {Notice that a possible strategy to assess the accuracy of
  this approximation, in the context of the present simulations, could consist
  in computing the exact maximum-entropy closing moments, from a converged
  approximated solution, and comparing them to the approximation
  itself.}\BibitemShut {Stop}%
\bibitem [{\citenamefont {Junk}\ and\ \citenamefont
  {Unterreiter}(2002)}]{junk2002maximum}%
  \BibitemOpen
  \bibfield  {author} {\bibinfo {author} {\bibfnamefont {M.}~\bibnamefont
  {Junk}}\ and\ \bibinfo {author} {\bibfnamefont {A.}~\bibnamefont
  {Unterreiter}},\ }\bibfield  {title} {\enquote {\bibinfo {title} {Maximum
  entropy moment systems and {Galilean} invariance},}\ }\href@noop {}
  {\bibfield  {journal} {\bibinfo  {journal} {Continuum Mechanics and
  Thermodynamics}\ }\textbf {\bibinfo {volume} {14}},\ \bibinfo {pages}
  {563--576} (\bibinfo {year} {2002})}\BibitemShut {NoStop}%
\bibitem [{\citenamefont {Taniguchi}\ and\ \citenamefont
  {Ruggeri}(2018)}]{taniguchi2018sub}%
  \BibitemOpen
  \bibfield  {author} {\bibinfo {author} {\bibfnamefont {S.}~\bibnamefont
  {Taniguchi}}\ and\ \bibinfo {author} {\bibfnamefont {T.}~\bibnamefont
  {Ruggeri}},\ }\bibfield  {title} {\enquote {\bibinfo {title} {On the
  sub-shock formation in extended thermodynamics},}\ }\href@noop {} {\bibfield
  {journal} {\bibinfo  {journal} {International Journal of Non-Linear
  Mechanics}\ }\textbf {\bibinfo {volume} {99}},\ \bibinfo {pages} {69--78}
  (\bibinfo {year} {2018})}\BibitemShut {NoStop}%
\bibitem [{\citenamefont {Parodi}\ \emph {et~al.}(2021)\citenamefont {Parodi},
  \citenamefont {Boccelli}, \citenamefont {Bariselli}, \citenamefont
  {Le~Quang}, \citenamefont {Lapenta},\ and\ \citenamefont
  {Magin}}]{parodi2021pic}%
  \BibitemOpen
  \bibfield  {author} {\bibinfo {author} {\bibfnamefont {P.}~\bibnamefont
  {Parodi}}, \bibinfo {author} {\bibfnamefont {S.}~\bibnamefont {Boccelli}},
  \bibinfo {author} {\bibfnamefont {F.}~\bibnamefont {Bariselli}}, \bibinfo
  {author} {\bibfnamefont {D.}~\bibnamefont {Le~Quang}}, \bibinfo {author}
  {\bibfnamefont {G.}~\bibnamefont {Lapenta}},\ and\ \bibinfo {author}
  {\bibfnamefont {T.}~\bibnamefont {Magin}},\ }\bibfield  {title} {\enquote
  {\bibinfo {title} {{PIC-MCC} characterization of expanding plasma plumes for
  a low-density hypersonic aerodynamics facility},}\ }in\ \href@noop {} {\emph
  {\bibinfo {booktitle} {APS Annual Gaseous Electronics Meeting Abstracts}}}\
  (\bibinfo {year} {2021})\ pp.\ \bibinfo {pages} {ET44--005}\BibitemShut
  {NoStop}%
\bibitem [{\citenamefont {Toro}\ \emph {et~al.}(2020)\citenamefont {Toro},
  \citenamefont {Saggiorato}, \citenamefont {Tokareva},\ and\ \citenamefont
  {Hidalgo}}]{toro2020low}%
  \BibitemOpen
  \bibfield  {author} {\bibinfo {author} {\bibfnamefont {E.~F.}\ \bibnamefont
  {Toro}}, \bibinfo {author} {\bibfnamefont {B.}~\bibnamefont {Saggiorato}},
  \bibinfo {author} {\bibfnamefont {S.}~\bibnamefont {Tokareva}},\ and\
  \bibinfo {author} {\bibfnamefont {A.}~\bibnamefont {Hidalgo}},\ }\bibfield
  {title} {\enquote {\bibinfo {title} {Low-dissipation centred schemes for
  hyperbolic equations in conservative and non-conservative form},}\
  }\href@noop {} {\bibfield  {journal} {\bibinfo  {journal} {Journal of
  Computational Physics}\ }\textbf {\bibinfo {volume} {416}},\ \bibinfo {pages}
  {109545} (\bibinfo {year} {2020})}\BibitemShut {NoStop}%
\bibitem [{\citenamefont {Van~Leer}(1979)}]{van1979towards}%
  \BibitemOpen
  \bibfield  {author} {\bibinfo {author} {\bibfnamefont {B.}~\bibnamefont
  {Van~Leer}},\ }\bibfield  {title} {\enquote {\bibinfo {title} {Towards the
  ultimate conservative difference scheme. v. a second-order sequel to
  godunov's method},}\ }\href@noop {} {\bibfield  {journal} {\bibinfo
  {journal} {Journal of computational Physics}\ }\textbf {\bibinfo {volume}
  {32}},\ \bibinfo {pages} {101--136} (\bibinfo {year} {1979})}\BibitemShut
  {NoStop}%
\bibitem [{\citenamefont {Toro}(2013)}]{toro2013riemann}%
  \BibitemOpen
  \bibfield  {author} {\bibinfo {author} {\bibfnamefont {E.~F.}\ \bibnamefont
  {Toro}},\ }\href@noop {} {\emph {\bibinfo {title} {Riemann solvers and
  numerical methods for fluid dynamics: a practical introduction}}}\ (\bibinfo
  {publisher} {Springer Science \& Business Media},\ \bibinfo {year}
  {2013})\BibitemShut {NoStop}%
\bibitem [{\citenamefont {{Boccelli, Stefano}}(view)}]{hyper2Dgithub}%
  \BibitemOpen
  \bibfield  {author} {\bibinfo {author} {\bibnamefont {{Boccelli, Stefano}}},\
  }\href@noop {} {\enquote {\bibinfo {title} {Hyper2d: a finite-volume solver
  for hyperbolic equations and non-equilibrium flows},}\ } (\bibinfo {year}
  {Under Review})\BibitemShut {NoStop}%
\bibitem [{\citenamefont {Rovenskaya}(2022)}]{rovenskaya2022subsonic}%
  \BibitemOpen
  \bibfield  {author} {\bibinfo {author} {\bibfnamefont {O.~I.}\ \bibnamefont
  {Rovenskaya}},\ }\bibfield  {title} {\enquote {\bibinfo {title} {Subsonic
  rarefied gas flow over a rectangular cylinder},}\ }\href@noop {} {\bibfield
  {journal} {\bibinfo  {journal} {Computational Mathematics and Mathematical
  Physics}\ }\textbf {\bibinfo {volume} {62}},\ \bibinfo {pages} {1928--1941}
  (\bibinfo {year} {2022})}\BibitemShut {NoStop}%
\bibitem [{\citenamefont {Muntz}\ and\ \citenamefont
  {Harnett}(1969)}]{muntz1969molecular}%
  \BibitemOpen
  \bibfield  {author} {\bibinfo {author} {\bibfnamefont {E.}~\bibnamefont
  {Muntz}}\ and\ \bibinfo {author} {\bibfnamefont {L.}~\bibnamefont
  {Harnett}},\ }\bibfield  {title} {\enquote {\bibinfo {title} {Molecular
  velocity distribution function measurements in a normal shock wave},}\
  }\href@noop {} {\bibfield  {journal} {\bibinfo  {journal} {The physics of
  fluids}\ }\textbf {\bibinfo {volume} {12}},\ \bibinfo {pages} {2027--2035}
  (\bibinfo {year} {1969})}\BibitemShut {NoStop}%
\bibitem [{\citenamefont {McDonald}(2011)}]{mcdonald2011extended}%
  \BibitemOpen
  \bibfield  {author} {\bibinfo {author} {\bibfnamefont {J.~G.}\ \bibnamefont
  {McDonald}},\ }\href@noop {} {\emph {\bibinfo {title} {Extended fluid-dynamic
  modelling for numerical solution of micro-scale flows}}}\ (\bibinfo
  {publisher} {University of Toronto},\ \bibinfo {year} {2011})\BibitemShut
  {NoStop}%
\bibitem [{\citenamefont {McDonald}\ and\ \citenamefont
  {Groth}(2008)}]{mcdonald2008extended}%
  \BibitemOpen
  \bibfield  {author} {\bibinfo {author} {\bibfnamefont {J.~G.}\ \bibnamefont
  {McDonald}}\ and\ \bibinfo {author} {\bibfnamefont {C.}~\bibnamefont
  {Groth}},\ }\bibfield  {title} {\enquote {\bibinfo {title} {Extended fluid
  dynamic model for micron-scale flows based on gaussian moment closure},}\
  }in\ \href@noop {} {\emph {\bibinfo {booktitle} {46th AIAA Aerospace Sciences
  Meeting and Exhibit}}}\ (\bibinfo {year} {2008})\ p.\ \bibinfo {pages}
  {691}\BibitemShut {NoStop}%
\bibitem [{\citenamefont {Holtz}\ and\ \citenamefont
  {Muntz}(1983)}]{holtz1983molecular}%
  \BibitemOpen
  \bibfield  {author} {\bibinfo {author} {\bibfnamefont {T.}~\bibnamefont
  {Holtz}}\ and\ \bibinfo {author} {\bibfnamefont {E.}~\bibnamefont {Muntz}},\
  }\bibfield  {title} {\enquote {\bibinfo {title} {Molecular velocity
  distribution functions in an argon normal shock wave at {Mach} number 7},}\
  }\href@noop {} {\bibfield  {journal} {\bibinfo  {journal} {The Physics of
  fluids}\ }\textbf {\bibinfo {volume} {26}},\ \bibinfo {pages} {2425--2436}
  (\bibinfo {year} {1983})}\BibitemShut {NoStop}%
\bibitem [{\citenamefont {Munaf{\`o}}\ \emph {et~al.}(2014)\citenamefont
  {Munaf{\`o}}, \citenamefont {Haack}, \citenamefont {Gamba},\ and\
  \citenamefont {Magin}}]{munafo2014spectral}%
  \BibitemOpen
  \bibfield  {author} {\bibinfo {author} {\bibfnamefont {A.}~\bibnamefont
  {Munaf{\`o}}}, \bibinfo {author} {\bibfnamefont {J.~R.}\ \bibnamefont
  {Haack}}, \bibinfo {author} {\bibfnamefont {I.~M.}\ \bibnamefont {Gamba}},\
  and\ \bibinfo {author} {\bibfnamefont {T.~E.}\ \bibnamefont {Magin}},\
  }\bibfield  {title} {\enquote {\bibinfo {title} {A spectral-{Lagrangian}
  {Boltzmann} solver for a multi-energy level gas},}\ }\href@noop {} {\bibfield
   {journal} {\bibinfo  {journal} {Journal of Computational Physics}\ }\textbf
  {\bibinfo {volume} {264}},\ \bibinfo {pages} {152--176} (\bibinfo {year}
  {2014})}\BibitemShut {NoStop}%
\bibitem [{\citenamefont {Bariselli}\ \emph {et~al.}(2018)\citenamefont
  {Bariselli}, \citenamefont {Boccelli}, \citenamefont {Magin}, \citenamefont
  {Frezzotti},\ and\ \citenamefont {Hubin}}]{bariselli2018aerothermodynamic}%
  \BibitemOpen
  \bibfield  {author} {\bibinfo {author} {\bibfnamefont {F.}~\bibnamefont
  {Bariselli}}, \bibinfo {author} {\bibfnamefont {S.}~\bibnamefont {Boccelli}},
  \bibinfo {author} {\bibfnamefont {T.}~\bibnamefont {Magin}}, \bibinfo
  {author} {\bibfnamefont {A.}~\bibnamefont {Frezzotti}},\ and\ \bibinfo
  {author} {\bibfnamefont {A.}~\bibnamefont {Hubin}},\ }\bibfield  {title}
  {\enquote {\bibinfo {title} {Aerothermodynamic modelling of meteor entry
  flows in the rarefied regime},}\ }in\ \href@noop {} {\emph {\bibinfo
  {booktitle} {2018 Joint Thermophysics and Heat Transfer Conference}}}\
  (\bibinfo {year} {2018})\ p.\ \bibinfo {pages} {4180}\BibitemShut {NoStop}%
\bibitem [{\citenamefont {Forgues}\ and\ \citenamefont
  {McDonald}(2019)}]{forgues2019higher}%
  \BibitemOpen
  \bibfield  {author} {\bibinfo {author} {\bibfnamefont {F.}~\bibnamefont
  {Forgues}}\ and\ \bibinfo {author} {\bibfnamefont {J.~G.}\ \bibnamefont
  {McDonald}},\ }\bibfield  {title} {\enquote {\bibinfo {title} {Higher-order
  moment models for laminar multiphase flows with accurate particle-stream
  crossing},}\ }\href@noop {} {\bibfield  {journal} {\bibinfo  {journal}
  {International Journal of Multiphase Flow}\ }\textbf {\bibinfo {volume}
  {114}},\ \bibinfo {pages} {28--38} (\bibinfo {year} {2019})}\BibitemShut
  {NoStop}%
\bibitem [{\citenamefont {Desjardins}, \citenamefont {Fox},\ and\ \citenamefont
  {Villedieu}(2008)}]{desjardins2008quadrature}%
  \BibitemOpen
  \bibfield  {author} {\bibinfo {author} {\bibfnamefont {O.}~\bibnamefont
  {Desjardins}}, \bibinfo {author} {\bibfnamefont {R.~O.}\ \bibnamefont
  {Fox}},\ and\ \bibinfo {author} {\bibfnamefont {P.}~\bibnamefont
  {Villedieu}},\ }\bibfield  {title} {\enquote {\bibinfo {title} {A
  quadrature-based moment method for dilute fluid-particle flows},}\
  }\href@noop {} {\bibfield  {journal} {\bibinfo  {journal} {Journal of
  Computational Physics}\ }\textbf {\bibinfo {volume} {227}},\ \bibinfo {pages}
  {2514--2539} (\bibinfo {year} {2008})}\BibitemShut {NoStop}%
\bibitem [{\citenamefont {Patel}, \citenamefont {Desjardins},\ and\
  \citenamefont {Fox}(2019)}]{patel2019three}%
  \BibitemOpen
  \bibfield  {author} {\bibinfo {author} {\bibfnamefont {R.~G.}\ \bibnamefont
  {Patel}}, \bibinfo {author} {\bibfnamefont {O.}~\bibnamefont {Desjardins}},\
  and\ \bibinfo {author} {\bibfnamefont {R.~O.}\ \bibnamefont {Fox}},\
  }\bibfield  {title} {\enquote {\bibinfo {title} {Three-dimensional
  conditional hyperbolic quadrature method of moments},}\ }\href@noop {}
  {\bibfield  {journal} {\bibinfo  {journal} {Journal of Computational Physics:
  X}\ }\textbf {\bibinfo {volume} {1}},\ \bibinfo {pages} {100006} (\bibinfo
  {year} {2019})}\BibitemShut {NoStop}%
\bibitem [{\citenamefont {Giroux}\ and\ \citenamefont
  {McDonald}(2021)}]{giroux2021approximation}%
  \BibitemOpen
  \bibfield  {author} {\bibinfo {author} {\bibfnamefont {F.}~\bibnamefont
  {Giroux}}\ and\ \bibinfo {author} {\bibfnamefont {J.~G.}\ \bibnamefont
  {McDonald}},\ }\bibfield  {title} {\enquote {\bibinfo {title} {An
  approximation for the twenty-one-moment maximum-entropy model of rarefied gas
  dynamics},}\ }\href@noop {} {\bibfield  {journal} {\bibinfo  {journal}
  {International Journal of Computational Fluid Dynamics}\ }\textbf {\bibinfo
  {volume} {35}},\ \bibinfo {pages} {632--652} (\bibinfo {year}
  {2021})}\BibitemShut {NoStop}%
\end{thebibliography}

%

\end{document}